\documentclass
[superscriptaddress,secnumarabic,amssymb,amsmath,nobibnotes,aps,prd,showkeys,nofootinbib,nopacsnumber,onecolumn,12pt]{revtex4}%
\usepackage{graphicx}
\usepackage{bm}
\usepackage{amsmath}
\usepackage{amsfonts}
\usepackage{amssymb}%
\providecommand{\U}[1]{\protect\rule{.1in}{.1in}}

\newcommand{\be}{\begin{equation}}
\newcommand{\ee}{\end{equation}}

\newcommand{\mincir}{\raise
-3.truept\hbox{\rlap{\hbox{$\sim$}}\raise4.truept\hbox{$<$}\ }}
\newcommand{\magcir}{\raise
-3.truept\hbox{\rlap{\hbox{$\sim$}}\raise4.truept\hbox{$>$}\ }}

\begin{document}
\title{Dynamical analysis in Chameleon dark energy}
\author{Andronikos Paliathanasis}
\email{anpaliat@phys.uoa.gr}
\affiliation{Institute of Systems Science, Durban University of Technology, Durban 4000,
South Africa}
\affiliation{Departamento de Matem\'{a}ticas, Universidad Cat\'{o}lica del Norte, Avda.
Angamos 0610, Casilla 1280 Antofagasta, Chile}

\begin{abstract}
We present a detailed analysis of the phase-space for the field equations in
scalar field cosmology with a chameleon cosmology in a spatially flat
Friedmann--Lema\^{\i}tre--Robertson--Walker spacetime. For the matter source
we assume that it is an ideal gas with a constant equation of state parameter,
while for the scalar field potential and the coupling function of the
chameleon mechanism we consider four different sets which provide four
different models. We consider the $H$-normalization approach and we write the
field equations with the help of dimensionless variables. The asymptotic
solutions are determined from where we find that the theory can describe the
main eras of cosmological history and evolution. Future attractors which
describe acceleration exist, however we found past acceleration solutions
related to the inflationary era, as also the radiation epoch and the matter
dominated eras are provided by the dynamics. We conclude that the Chameleon
dark energy model can be used as a unified model for the elements which
contribute to the dark sector of the universe.

\end{abstract}
\keywords{Chameleon gravity; Scalar field; Cosmology;\ Phase-space analysis; Dynamical analysis}\date{\today}
\maketitle

\section{Introduction}

\label{sec1}

The recent cosmological data \cite{rr1,Teg,Kowal,Komatsu,suzuki11} indicate
that the universe is in an acceleration phase. Dark energy attributes the
late-time acceleration of the universe \cite{jo}. Indeed, dark energy is a
fluid source with negative pressure such as to provide anti-gravitating
effects. The physical origin and the nature of dark energy are still unknown;
however, there are various proposals in the literature, see for instance
\cite{de01,de02,de03,de04} and references therein.

Scalar field models are of special interest in gravitational theories for
which they play an important role in the explanation of cosmological
observations. Scalar fields introduce new degrees of freedom in the
gravitational field equations. The new dynamical variables provide a mechanism
which can drive the evolution of the physical parameters as provided by the
cosmological data. The simplest scalar field model is the quintessence theory
\cite{ratra}. In this theory the equation of state parameter has as lower
bound the value $-1$ which is the limit of the cosmological constant and an
upper bound of $+1$ which describes a stiff fluid. The quintessence scalar
field satisfies the null energy condition, the weak energy condition and the
dominant energy condition. However, because the fluid pressure component can
be negative and provide acceleration, the strong energy condition for the
quintessence can be violated. The dynamics of the quintessence cosmological
model with an exponential potential was investigated in detail in \cite{q5}.
The dynamical analysis provides the cosmological model and it admits a stiff
fluid solution, a scaling solution which can describe acceleration, a matter
dominated solution and a tracking solution where the scalar field has the same
physical behaviour as the matter source. In the absence of a matter source,
the stiff fluid and the scaling solutions exist for the exponential potential.
On the hand, for a power-law potential function the tracking solution does not
exist, but then a de Sitter solution appears where the scalar field reaches
the limit of the cosmological constant.

A scalar field similar to that of the quintessence theory has been used to
describe the inflaton mechanism \cite{inf2} responsible for the inflation
which has been introduced to solve the horizon and the isotropization problems
\cite{inf1}. Last but not least, scalar fields can attribute the higher-order
degrees of freedom provided by the modified theories of gravity, for instance
the quadratic inflationary model belongs to the $f\left(  R\right)  $-gravity
which after the application of a Lagrange multiplier and a conformal
transformation is equivalent to a quintessence scalar field theory
\cite{star1}. Because of the importance of the quintessence theory there is a
plethora of studies in the literature which investigate different functional
forms of the scalar field potential and derive new analytic and exact
solutions \cite{qq1,qq2,qq3,qq4,qq5,qq6,qq7,qq8,qq9,qq10}. Quintessence is a
simple dark energy which, however, cannot solve various puzzles such as the
cosmological tensions \cite{ht1}.\ Thus, various scalar fields have been
proposed in the literature, such as phantom scalar field models \cite{q14,q15}%
, Brans-Dicke and scalar-tensor theories \cite{Brans,sf9}, Galileons
\cite{gal1,gal2}, k-essence \cite{q23}, tachyons \cite{tach,tch}, multi-scalar
field models \cite{q22,q24,ml1} and others.

A geometric mechanism to introduce a minimally coupled to gravity scalar field
in the Einstein-Hilbert Action Integral of General Relativity is the Weyl
theory and specifically the Weyl Integrable Spacetime (WIS) \cite{salim96}.
This theory is a torsion-free theory embedded with two conformal related
metrics and the connection preserves the conformal structure. In WIS the
connection structure differs from the Levi-Civita connection by a scalar field
\cite{va5,va6,va7}. The novelty of WIS\ is that in the presence of a matter
source because of the conformal structure there appears a coupling between the
scalar field and the matter source. Hence, there exists an interaction between
the matter components of the gravitational theory and the mass of the scalar
field depends upon the energy density of the matter source \cite{hot,hot2}%
.\ Inspired by this property in \cite{ch1,ch2} there has been proposed a
chameleon mechanism which generalizes the coupling between the scalar field
and the matter source. In chameleon theory the WIS is only a particular case
for which the coupling function is an exponential. The limit of the WIS was
investigated recently in \cite{df1}, while the case for which the background
geometry has nonzero spatial curvature was considered in \cite{df2}. A
power-law coupling was investigated in \cite{df3}; it was found that this
cosmological model for power-law potential has a behaviour very close to that
of $\Lambda$CDM theory, but in low redshifts the model can enhance the growth
of the linear perturbations. A tachyonic-like chameleon model was investigated
in \cite{df4} while some other generalizations can be found in
\cite{df5,df6,df7,df8}.

In this piece of work we investigate the dynamical evolution and the
asymptotic solutions of chameleon cosmology in \ the context of a homogeneous
and isotropic spatially flat Friedmann--Lema\^{\i}tre--Robertson--Walker
spacetime. For the matter source we consider an ideal gas and we assume
various sets for the free functions of the theory; they are the scalar field
potential and the coupling function. We make use of the H-normalization
approach \cite{q5} and we determine the stationary points for the field
equations and we calculate their stability properties. We wish to answer to
the question if the chameleon dark energy model can explain the main eras of
the cosmological history. Furthermore we make conclusions for the initial
value problem in this cosmological theory. The mathematical methods that we
apply in this work have been successful for the classification of various
cosmological models \cite{dn1,dn2,dn3,dn4,dn5}. The structure of the paper is
as follows.

In Section \ref{sec2} we introduce the cosmological theory of our
consideration which is that of General Relativity with a scalar field and a
chameleon mechanism, that is, the scalar field is coupled to a perfect fluid.
Furthermore, we assume that the universe is described by the spatially flat
Friedmann--Lema\^{\i}tre--Robertson--Walker (FLRW) geometry. The field
equations are of second-order with dynamical variables the scale factor, the
scalar field and the energy density of the perfect fluid which is assumed to
be an ideal gas with constant equation of state parameter. Over and above, the
dynamical evolution of the physical \ variables depends upon two unknown
functions, the scalar field potential $V\left(  \phi\right)  $ and the
coupling function, $f\left(  \phi\right) , $ responsible for the chameleon
mechanism. In order to investigate the dynamics of the cosmological field
equations we consider new dimensionless variables. Thus in Section \ref{sec3}
we write the field equations in the equivalent form of an
algebraic-differential system. We consider four-different sets for the
potential and the coupling functions and we determine the stationary points
and their stability properties as the physical properties of the asymptotic
solutions at the stationary points. In Section \ref{sec4} we study the case
for which $V\left(  \phi\right)  =V_{0}e^{\lambda_{0}\phi}$~and~$f\left(
\phi\right)  =f_{0}e^{\sigma_{0}\phi}$. The cosmological constant term is
introduced in Section \ref{sec5} in which we select $V\left(  \phi\right)
=V_{0}e^{\lambda_{0}\phi}+\Lambda$ and $f\left(  \phi\right)  =f_{0}%
e^{\sigma_{0}\phi}$. The dynamical analysis of the field equations for the
functions $V\left(  \phi\right)  =V_{0}e^{\lambda_{0}\phi}$ and $f\left(
\phi\right)  =f_{0}e^{\sigma_{0}\phi}+\Lambda$ is performed in Section
\ref{sec6}. For the fourth model of our analysis we select $V\left(
\phi\right)  =V_{0}e^{\lambda_{0}\phi}+\Lambda$ and $f\left(  \phi\right)
=f_{0}\left(  V_{0}e^{\sigma_{0}\phi}+\Lambda\right)  ^{p}$, $p\neq0$ and its
phase-space analysis is presented in Section \ref{sec7}. Finally, in Section
\ref{sec8} we summarize our results and we draw our conclusions.

\section{Chameleon dark energy}

\label{sec2}

The gravitational Action Integral of Chameleon dark energy is \cite{ch1}
\begin{equation}
S=\int\sqrt{-g}d^{4}x\left(  R-\frac{1}{2}g^{\mu\kappa}\nabla_{\mu}\phi\left(
x^{\nu}\right)  \nabla_{\kappa}\phi\left(  x^{\nu}\right)  -V\left(
\phi\left(  x^{\nu}\right)  \right)  -f\left(  \phi\right)  L_{m}\left(
x^{\nu}\right)  \right)  , \label{ac.01}%
\end{equation}
where $R$ is the Ricci scalar for the four-dimensional background Riemannian
physical space with metric tensor $g_{\mu\nu}$; $\phi\left(  x^{\mu}\right)  $
is a scalar with potential function $V\left(  \phi\left(  x^{\mu}\right)
\right)  $ and $L_{m}\left(  x^{\mu}\right)  $ is a Lagrangian function for
the matter source. For this we assume an ideal gas with energy density
$\rho_{m}$ and pressure component $p_{m}$ and an equation of state parameter
$w_{m}$, that is $p_{m}=w_{m}\rho_{m}.$ Hence, the Lagrangian for the matter
source is expressed as $L_{m}\left(  x^{\mu}\right)  $ $\simeq\rho_{m}%
~$\cite{lan1}. The coupling function $f\left(  \phi\right)  $ between the
scalar field and the matter source describes the chameleon mechanism.

The gravitational field equations are%
\begin{equation}
G_{\mu\nu}=T_{\mu\nu}^{eff}, \label{ac.02}%
\end{equation}
where $G_{\mu\nu}$ is the Einstein tensor and $T_{\mu\nu}^{eff}$ is the
effective energy-momentum tensor, that is, $T_{\mu\nu}^{eff}=T_{\mu\nu}^{\phi
}+f\left(  \phi\right)  T_{\mu\nu}^{m}$.\ The latter are defined as
\begin{equation}
T_{\mu\nu}^{\phi}=\nabla_{\mu}\phi\nabla_{\nu}\phi-g_{\mu\nu}\left(  \frac
{1}{2}g^{\kappa\zeta}\nabla_{\kappa}\phi\nabla_{\zeta}\phi+V\left(
\phi\right)  \right)  ,
\end{equation}
and
\begin{equation}
T_{\mu\nu}^{m}=\left(  \rho_{m}+p_{m}\right)  u_{\mu}u_{\nu}+p_{m}g_{\mu\nu},
\end{equation}
where $u_{\mu}$ is the co-moving observer with $g_{\mu\nu}u^{\mu}u^{\nu}=-1$.

The equation of motion for the matter source is $\nabla_{\nu}T^{eff~~\mu\nu
}=0$, i.e.
\begin{equation}
\nabla_{\nu}\left(  T^{\phi~\mu\nu}+f\left(  \phi\right)  T^{m~\mu\nu}\right)
=0,
\end{equation}
or%
\begin{equation}
-g^{\mu\nu}\nabla_{\mu}\nabla_{\nu}\phi+V\left(  \phi\right)  +\nabla_{\mu
}\left(  f\left(  \phi\right)  \rho_{m}\right)  u^{\mu}+f\left(  \phi\right)
\left(  \rho_{m}+p_{m}\right)  \nabla_{\mu}u^{\mu}=0. \label{sd1}%
\end{equation}

Equivalently we can write the following equations \cite{ch1}%
\begin{equation}
-g^{\mu\nu}\nabla_{\mu}\nabla_{\nu}\phi+V\left(  \phi\right)  +\left(
1+\alpha\right)  \rho_{m}\nabla_{\mu}f\left(  \phi\right)  u^{\mu}=0,
\label{sd12}%
\end{equation}%
\begin{equation}
\nabla_{\mu}\left(  \rho_{m}\right)  u^{\mu}+\left(  \rho_{m}+p_{m}\right)
\nabla_{\mu}u^{\mu}-\alpha\rho_{m}\nabla_{\mu}\ln\left(  f\left(  \phi\right)
\right)  u^{\mu}=0. \label{sd14}%
\end{equation}
The parameter, $\alpha$, is an arbitrary parameter different from zero and
minus one. It has been introduced only in order to write equation (\ref{sd1})
as a system of two equations.

For a spatially flat FLRW line element
\begin{equation}
ds^{2}=-dt^{2}+a^{2}\left(  t\right)  \left(  dx^{2}+dy^{2}+dz^{2}\right)
\end{equation}
in which $a\left(  t\right)  $ is the scalar factor and $H=\frac{\dot{a}}{a}$
is the Hubble function, with $\dot{a}=\frac{da}{dt}$, the gravitational field
equations (\ref{ac.01}) are written as follows \cite{ch1}
\begin{equation}
3H^{2}=\frac{1}{2}\dot{\phi}^{2}+V\left(  \phi\right)  +\rho_{m}f\left(
\phi\right)  \label{eq.01}%
\end{equation}
and
\begin{equation}
2\dot{H}+3H^{2}=-\left(  \frac{1}{2}\dot{\phi}^{2}-V\left(  \phi\right)
+f\left(  \phi\right)  p_{m}\right)  \label{eq.02}%
\end{equation}
with the equations of motion%
\begin{equation}
\ddot{\phi}+3H\dot{\phi}+V_{,\phi}+\left(  1+\alpha\right)  f_{,\phi}\rho
_{m}=0, \label{eq.03}%
\end{equation}
and
\begin{equation}
\dot{\rho}_{m}+3H\left(  \rho_{m}+p_{m}\right)  -\alpha\dot{\phi}\left(  \ln
f\right)  _{,\phi}\rho_{m}=0. \label{eq.04}%
\end{equation}

We have assumed that the scalar field and the matter source inherit the
symmetries of the background space, that is, $\phi=\phi\left(  t\right)  $ and
$\rho_{m}=\rho_{m}\left(  t\right)  $. Moreover, for the matter source we
consider a constant equation of state parameter $w_{m}=\frac{p_{m}}{\rho_{m}}$
with $0\leq w_{m}<1$. In the limit $w_{m}=0$ the fluid is pressureless, while
for $w_{m}=1$ the fluid is a stiff fluid which can be described by a massless
scalar field. From the modified Klein-Gordon equation (\ref{eq.03}) we observe
that the mass of the scalar field depends upon the coupling function $f\left(
\phi\right)  $ and upon the energy density of the scalar field.

\section{Phase-space analysis}

\label{sec3}

We proceed to the analysis of the dynamics for the field equations
(\ref{eq.01})-(\ref{eq.04}). In particular we investigate the stationary
points of the phase-space and we study their stability properties.

We work in the $H$-normalization approach \cite{q5} and we define the
dimensionless variables and parameters%
\begin{equation}
\tau=\ln a,~x=\frac{\dot{\phi}}{\sqrt{6}H},~y=\sqrt{\frac{V\left(
\phi\right)  }{3H}},~\Omega_{m}=\frac{\rho_{m}f\left(  \phi\right)  }{3H^{2}},
\end{equation}%
\begin{equation}
\lambda=\frac{V_{,\phi}}{V}~,~\sigma=\frac{f_{,\phi}}{f}~,~\Gamma\left(
\lambda\right)  =\frac{V_{,\phi\phi}V}{\left(  V_{,\phi}\right)  ^{2}%
}~,~\Delta\left(  \lambda\right)  =\frac{f_{,\phi\phi}f}{\left(  f_{,\phi
}\right)  ^{2}}.
\end{equation}

In the new variables the field equations are expressed as the following
algebraic differential system of first-order differential equations%
\begin{equation}
\frac{dx}{d\tau}=\frac{1}{2}\left(  3x\left(  x^{2}-y^{2}-1+w_{m}\Omega
_{m}\right)  -\sqrt{6}\left(  \lambda y^{2}+\left(  1+\alpha\right)
\sigma\Omega_{m}\right)  \right)  , \label{de.01}%
\end{equation}%
\begin{equation}
\frac{dy}{d\tau}=\frac{1}{2}y\left(  3+\sqrt{6}\lambda x+3\left(  x^{2}%
-y^{2}+w_{m}\Omega_{m}\right)  \right)  , \label{de.02}%
\end{equation}%
\begin{equation}
\frac{d\Omega_{m}}{d\tau}=\Omega_{m}\left(  \sqrt{6}\left(  1+\alpha\right)
\sigma x+3\left(  x^{2}-y^{2}+w_{m}\left(  \Omega_{m}-1\right)  \right)
\right)  , \label{de.03}%
\end{equation}%
\begin{equation}
\frac{d\lambda}{d\tau}=\sqrt{6}x\lambda^{2}\left(  \Gamma\left(
\lambda\right)  -1\right)  , \label{de.04}%
\end{equation}%
\begin{equation}
\frac{d\sigma}{d\tau}=\sqrt{6}x\sigma^{2}\left(  \Delta\left(  \sigma\right)
-1\right)  , \label{de.05}%
\end{equation}
and from (\ref{eq.01}) we derive the algebraic constraint%
\begin{equation}
1-x^{2}-y^{2}-\Omega_{m}=0. \label{de.06}%
\end{equation}

We observe that not all the variables are independent; indeed, by definition
$\lambda=\lambda\left(  \phi\right)  $ and $\sigma=\sigma\left(  \phi\right)
$, which means that for arbitrary functions $V\left(  \phi\right)  $ and
$f\left(  \phi\right)  $ it follows that $\sigma=\sigma\left(  \lambda\right)
$ or $\lambda=\lambda\left(  \sigma\right)  $. Moreover, with the use of the
constraint equation (\ref{de.06}) the dimension of the dynamical system
(\ref{de.01})-(\ref{de.05}) has maximum value three.

Furthermore, we assume a positive coupling function, $f\left(  \phi\right)
\geq0$, from which it follows that $\Omega_{m}\geq0$. Thus from the constraint
equation (\ref{de.06}) parameters $x$ and $y$ are bounded on the
two-dimensional unitary disk, i.e. $x^{2}+y^{2}\leq1$, that is, $\left\vert
x\right\vert \leq1$ and $0<y\leq1$.

In the new-dimensionless variables the equation of state parameter for the
effective fluid, $w_{eff}=-1-\frac{2}{3}\frac{\dot{H}}{H^{2}}$, becomes%
\begin{equation}
w_{eff}\left(  x,y,\Omega_{m};w_{m}\right)  =x^{2}-y^{2}+w_{m}\Omega_{m}.
\end{equation}

Each stationary point, $P=x_{0}\left(  P\right)  $, where $x=\left(
x,y,\Omega_{m},\lambda,\sigma\right)  ^{T}~$of the dynamical system
(\ref{de.01})-(\ref{de.05}) describes an asymptotic solution for the
cosmological model with effective equation of state parameter $w_{eff}\left(
P\right)  $ and scale factor $a\left(  t\right)  =a_{0}t^{\frac{2}{3\left(
1+w_{eff}\left(  P\right)  \right)  }}$ for $w_{eff}\left(  P\right)
\neq-1\,;\,\ $and $a\left(  t\right)  =a_{0}e^{H_{0}t}$ for $w_{eff}\left(
P\right)  =-1$. At each point $P$ we determine the stability properties of the
asymptotic solution with the study of the eigenvalues of the linearised system
around the stationary point. Hence we can constrain the free parameters of the
model and the initial conditions in order to reconstruct the cosmological history.

Below we define the following four sets for the scalar field potential
$V\left(  \phi\right)  $ and the coupling function $f\left(  \phi\right)  $.
(A) $V\left(  \phi\right)  =V_{0}e^{\lambda_{0}\phi}$~and~$f\left(
\phi\right)  =f_{0}e^{\sigma_{0}\phi}$; (B) $V\left(  \phi\right)
=V_{0}e^{\lambda_{0}\phi}+\Lambda$ and $f\left(  \phi\right)  =f_{0}%
e^{\sigma_{0}\phi}$; (C) $V\left(  \phi\right)  =V_{0}e^{\lambda_{0}\phi}$ and
$f\left(  \phi\right)  =f_{0}e^{\sigma_{0}\phi}+\Lambda$; (D) $V\left(
\phi\right)  =V_{0}e^{\lambda_{0}\phi}+\Lambda$ and $f\left(  \phi\right)
=f_{0}\left(  V_{0}e^{\sigma_{0}\phi}+\Lambda\right)  ^{p}$. The exponential
potential has been widely studied in the literature for the description of the
early and late-time acceleration phases of the universe \cite{exp1,exp2},
while in \cite{exp3,exp4,exp5} the cosmological constant term has been
introduced into the potential function. As far as the interaction is
concerned, the exponential interaction is related to the WIS \cite{salim96}%
\thinspace. 

\section{Model A: $V\left(  \phi\right)  =V_{0}e^{\lambda_{0}\phi}$ and
$f\left(  \phi\right)  =f_{0}e^{\sigma_{0}\phi}$}

\label{sec4}

For the first model we consider the exponential scalar field potential
$V\left(  \phi\right)  =V_{0}e^{\lambda_{0}\phi}$ and exponential coupling
function $f\left(  \phi\right)  =f_{0}e^{\sigma_{0}\phi}$. For these
definitions it follows that the $\lambda$ and $\sigma$ parameters are always
constant, $\lambda=\lambda_{0}$ and $\sigma=\sigma_{0}$. Moreover with the
application of the constraint equation (\ref{de.06}) the dimension of the
dynamical system is reduced to two. We replace $\Omega_{m}$ from (\ref{de.06})
and we have the two first-order differential equations
\begin{equation}
\frac{dx}{d\tau}=\frac{1}{2}\left(  3x\left(  x^{2}-y^{2}-1+w_{m}\left(
1-x^{2}-y^{2}\right)  \right)  -\sqrt{6}\left(  \lambda y^{2}+\left(
1+\alpha\right)  \sigma\left(  1-x^{2}-y^{2}\right)  \right)  \right)
\label{de.07}%
\end{equation}
and
\begin{equation}
\frac{dy}{d\tau}=\frac{1}{2}y\left(  3+\sqrt{6}\lambda x+3\left(  x^{2}%
-y^{2}+w_{m}\left(  1-x^{2}-y^{2}\right)  \right)  \right)  .\label{de.08}%
\end{equation}

\begin{figure}[ptb]
\centering\includegraphics[width=1\textwidth]{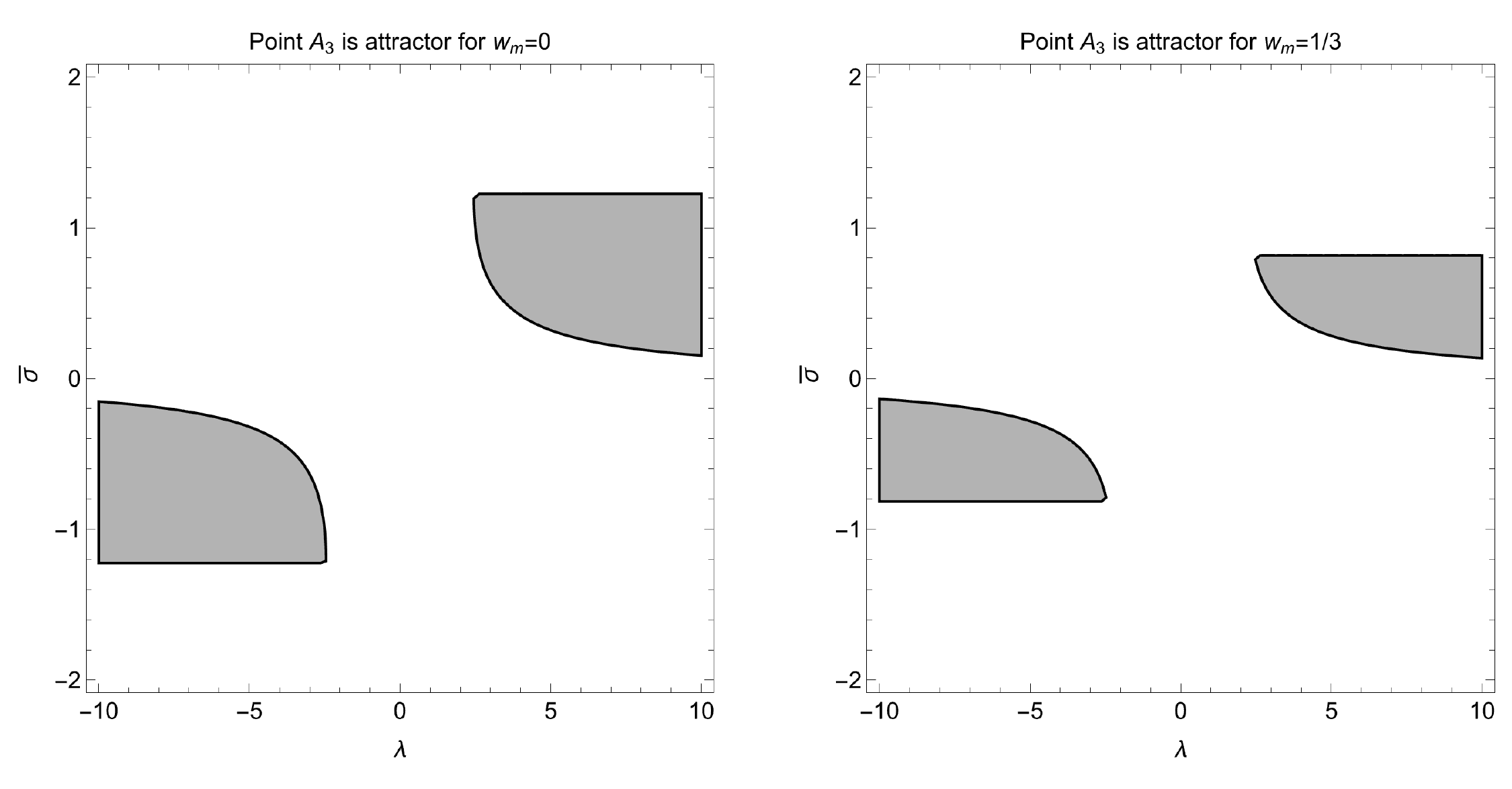}\caption{Region plot
on the space of the free parameters $\lambda$ and $\bar{\sigma}$ for $w_{m}=0$
(left fig.) and $w_{m}=\frac{1}{3}$ (right fig.) The shadowed areas correspond
the values of the parameters $\left(  \lambda,\bar{\sigma}\right)  $ in which
the stationary point $A_{3}$ is an attractor. }%
\label{mfig0}%
\end{figure}

The stationary points $A=\left(  x\left(  A\right)  ,y\left(  A\right)
\right)  $ of the dynamical system (\ref{de.07}), (\ref{de.08}) at the finite
regime are presented bellow.
\[
A_{1}^{\pm}=\left(  \pm1,0\right)  ,
\]
they describe scaling solutions in which the kinetic term of the scalar field
dominates in the cosmological fluid, that is, $\Omega_{m}\left(  A_{1}^{\pm
}\right)  =0$ and $w_{eff}\left(  A_{1}^{\pm}\right)  =1$. The eigenvalues of
the linearised system are derived to be $e_{1}\left(  A_{1}^{\pm}\right)
=\frac{1}{2}\left(  6\pm\sqrt{6}\lambda\right)  $ and $e_{2}\left(  A_{1}%
^{\pm}\right)  =3\left(  1-w_{m}\right)  \pm\sqrt{6}\bar{\sigma}$ with
$\bar{\sigma}=\left(  1+\alpha\right)  \sigma$. Hence, point $A_{1}$ is an
attractor for $\lambda<-\sqrt{6}$ and $\bar{\sigma}<-3\left(  1-w_{m}\right)
$, while point $A_{2}$ is an attractor for $\lambda>\sqrt{6}$ and~
$\bar{\sigma}>3\left(  1-w_{m}\right)  $.
\[
A_{2}=\left(  -\frac{\lambda}{\sqrt{6}},\sqrt{1-\frac{\lambda^{2}}{6}}\right)
\]
corresponds to a universe with $\Omega_{m}\left(  A_{2}\right)  =0$ and
$w_{eff}\left(  A_{2}\right)  =-1+\frac{\lambda^{2}}{3}$. The point is real
and physically acceptable for $\lambda^{2}\leq6$. When $\lambda=0$, that is
$V\left(  \phi\right)  =V_{0}$, the de Sitter universe is recovered. The
eigenvalues of the linearised system are~$e_{1}\left(  A_{2}\right)
=\frac{\lambda^{2}-6}{2}$ and $e_{2}\left(  A_{2}\right)  =-3\left(
1+w_{m}\right)  +\lambda\left(  \lambda-\bar{\sigma}\right)  $, from which we
conclude that the point is an attractor for $0\leq\lambda^{2}<6$ and
\thinspace$3\left(  w_{m}+1\right)  +\lambda^{2}>\lambda\bar{\sigma}$.
\[
A_{3}=\left(  -\sqrt{\frac{2}{3}}\frac{1+\alpha}{1-w_{m}}\sigma,0\right)
\]
describes a universe with $\Omega_{m}\left(  A_{3}\right)  =1-\frac
{2\bar{\sigma}^{2}}{3\left(  1-w_{m}\right)  ^{2}}$ and $w_{eff}\left(
A_{3}\right)  =\frac{2\bar{\sigma}^{2}}{3\left(  1-w_{m}\right)  }+w_{m}$. The
point is real and physically acceptable when~$\left\vert \sqrt{\frac{2}{3}%
}\frac{1+\alpha}{1-w_{m}}\sigma\right\vert \leq1$. The stationary points
describe an accelerated universe for $2\left(  \bar{\sigma}\right)
^{2}<\left(  2-3w_{m}\right)  -1$. We observe that this stationary point
cannot describe the de Sitter universe. The eigenvalues of the linearised
system are derived to be $e_{1}\left(  A_{3}\right)  =-\frac{3}{2}\left(
1-w_{m}\right)  +\frac{\left(  \bar{\sigma}\right)  ^{2}}{1-w_{m}}$ and
$e_{2}\left(  A_{3}\right)  =\frac{3}{2}\left(  1+w_{m}\right)  +\frac
{\bar{\sigma}\left(  \bar{\sigma}-\lambda\right)  }{1-w_{m}}$. Hence, the
stationary point is an attractor in the region presented in Fig. \ref{mfig0}.

\begin{figure}[ptbh]
\centering\includegraphics[width=1\textwidth]{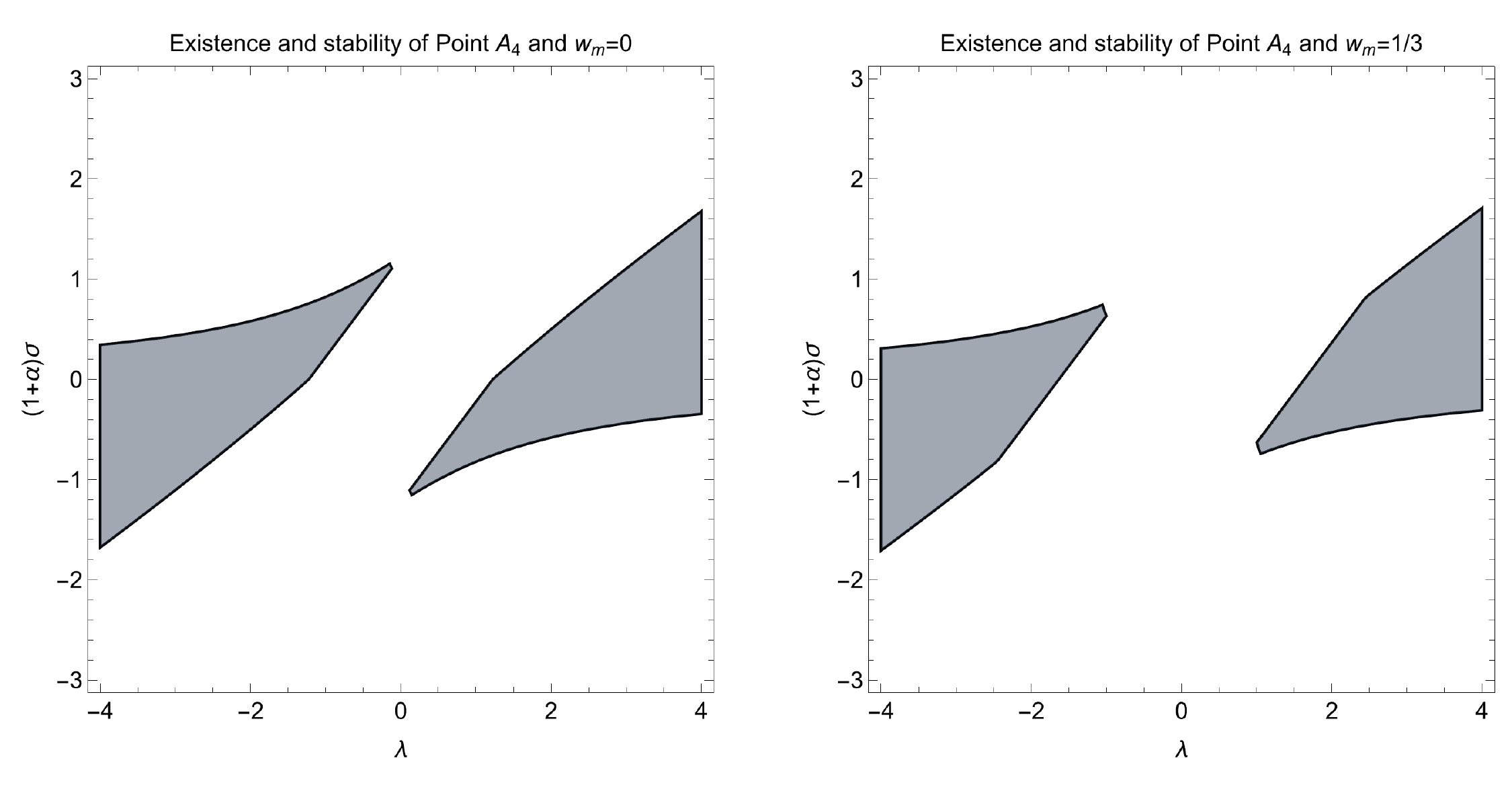}\caption{Region plot
on the space of the free parameters $\lambda$ and $\left(  1+\alpha\right)
\sigma$ for $w_{m}=0$ (left fig.) and $w_{m}=\frac{1}{3}$ (right fig.). The
shadowed areas correspond the values of the parameters $\left(  \lambda
,\left(  1+\alpha\right)  \sigma\right)  $ in which the stationary point
$A_{4}$ exists and when it is an attractor. We remark that point $A_{4}$ is
always an attractor when it exists}%
\label{mfig1}%
\end{figure}

\[
A_{4}=\left(  -\sqrt{\frac{3}{2}}\frac{1+w_{m}}{\lambda-\bar{\sigma}}%
,\frac{\sqrt{\frac{3\left(  w_{m}^{2}-1\right)  }{\bar{\sigma}-\lambda}%
+2\bar{\sigma}}}{\sqrt{2\lambda-2\bar{\sigma}}}\right)  .
\]
It has physical parameters $\Omega_{m}\left(  A_{4}\right)  =\frac
{\lambda\left(  \bar{\sigma}-\lambda_{0}\right)  -3\left(  1+w_{m}\right)
}{\left(  \lambda-\bar{\sigma}\right)  ^{2}}$ and $w_{eff}\left(
A_{4}\right)  =\frac{\bar{\sigma}+w_{m}\lambda}{\lambda-\bar{\sigma}}$. The
eigenvalues of the linearised system are $e_{1,2}\left(  A_{4}\right)
=\frac{1}{4}\left(  -3\left(  1-w_{m}\right)  +\frac{3\bar{\sigma}\left(
1+w_{m}\right)  }{\lambda-\bar{\sigma}}+\frac{\sqrt{X_{1}+X_{2}}}{\lambda
-\bar{\sigma}}\right)  $ with $X_{1}=12\bar{\sigma}\left(  \lambda\left(
4\lambda^{2}-9-3w_{m}\left(  3+2w_{m}\right)  \right)  +\bar{\sigma}\left(
15+12w_{m}-8\lambda^{2}+4\lambda\bar{\sigma}\right)  \right)  $ and
$X_{2}=9\left(  w_{m}-1\right)  \left(  \left(  7+9w_{m}\right)  \lambda
^{2}-24\left(  1+w_{m}^{2}\right)  \right)  $. In Fig. \ref{mfig1} we present
the regions in the space $\left\{  \lambda,\bar{\sigma}\right\}  $ in which
the stationary point $A_{4}$ is physically acceptable and when it is an attractor.

\begin{figure}[ptbh]
\centering\includegraphics[width=1\textwidth]{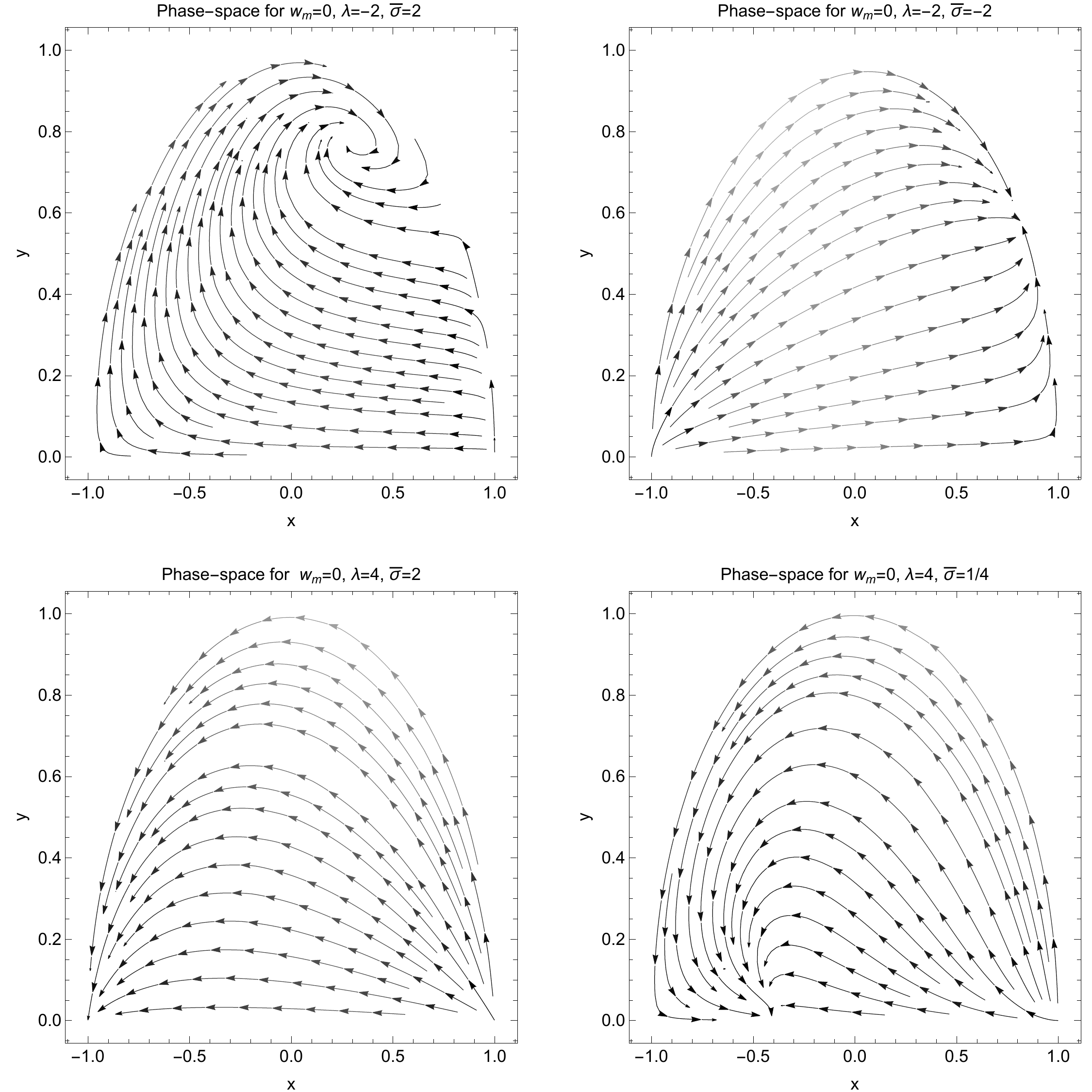}\caption{Phase-space
portraits of the dynamical system (\ref{de.07}), (\ref{de.08}) of Model A for
$w_{m}=0$. The free parameters have been selected such that the four
stationary points appear as attractors for different values of the free
parameters. For $\left(  \lambda,\bar{\sigma}\right)  =\left(  -2,2\right)  $,
point $A_{4}$ is the unique attractor; for $\left(  \lambda,\bar{\sigma
}\right)  =\left(  -2,-2\right)  $, point $A_{3}$ is the attractor of the
dynamical system. Moreover for $\left(  \lambda,\bar{\sigma}\right)  =\left(
4,2\right)  $ point $A_{1}^{-}$ is an attractor while for $\left(
\lambda,\bar{\sigma}\right)  =\left(  4,\frac{1}{4}\right)  $ point $A_{3}$ is
the attractor of the dynamical system. }%
\label{mfig2}%
\end{figure}

\begin{figure}[ptbh]
\centering\includegraphics[width=1\textwidth]{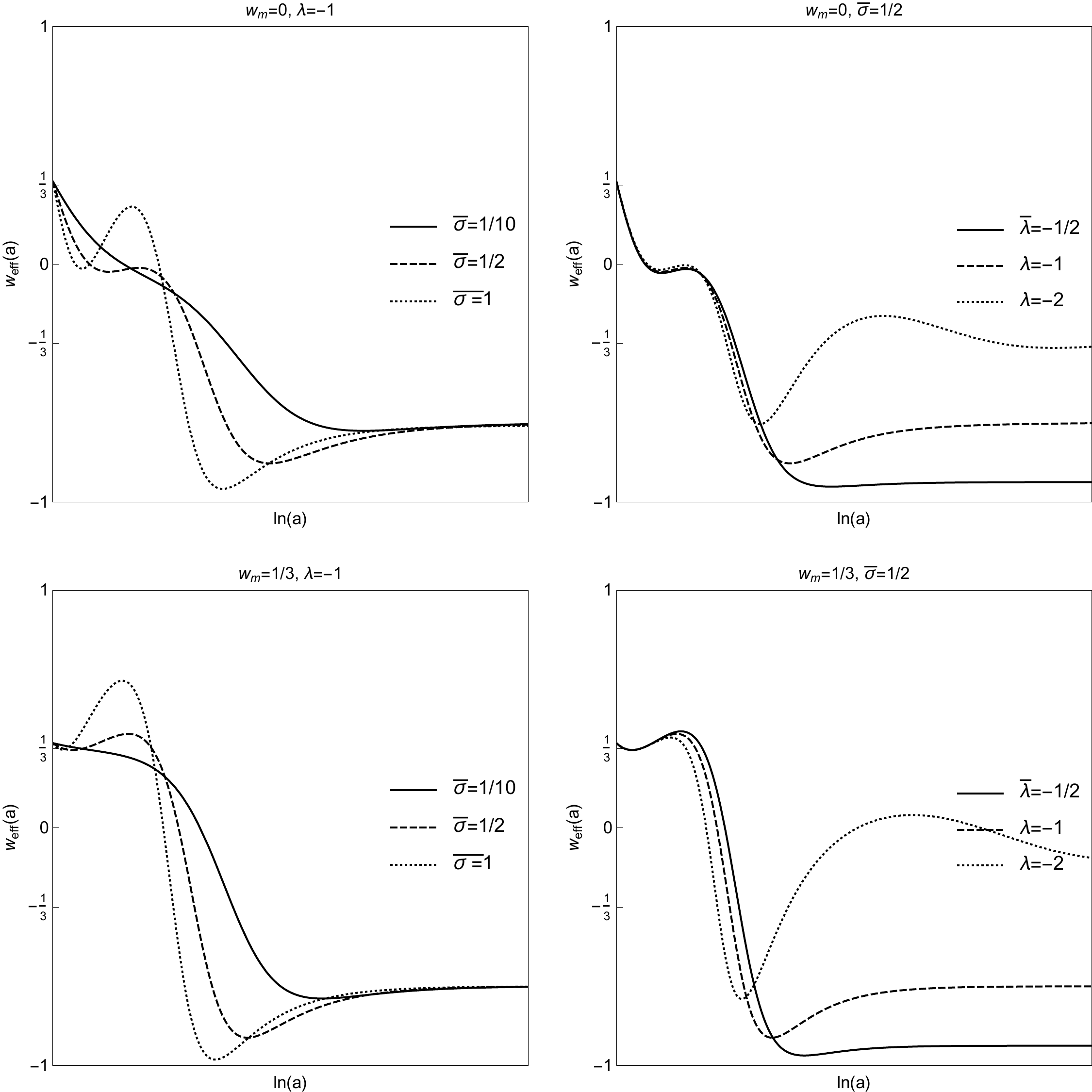}\caption{Qualitative
evolution for the effective equation of state parameter $w_{eff}\left(
a\right)  $ as it is given by the numerical solution of the dynamical system
(\ref{de.07}), (\ref{de.08}) of Model A for different values of the free
parameters. }%
\label{mfig3}%
\end{figure}

\begin{figure}[ptbh]
\centering\includegraphics[width=1\textwidth]{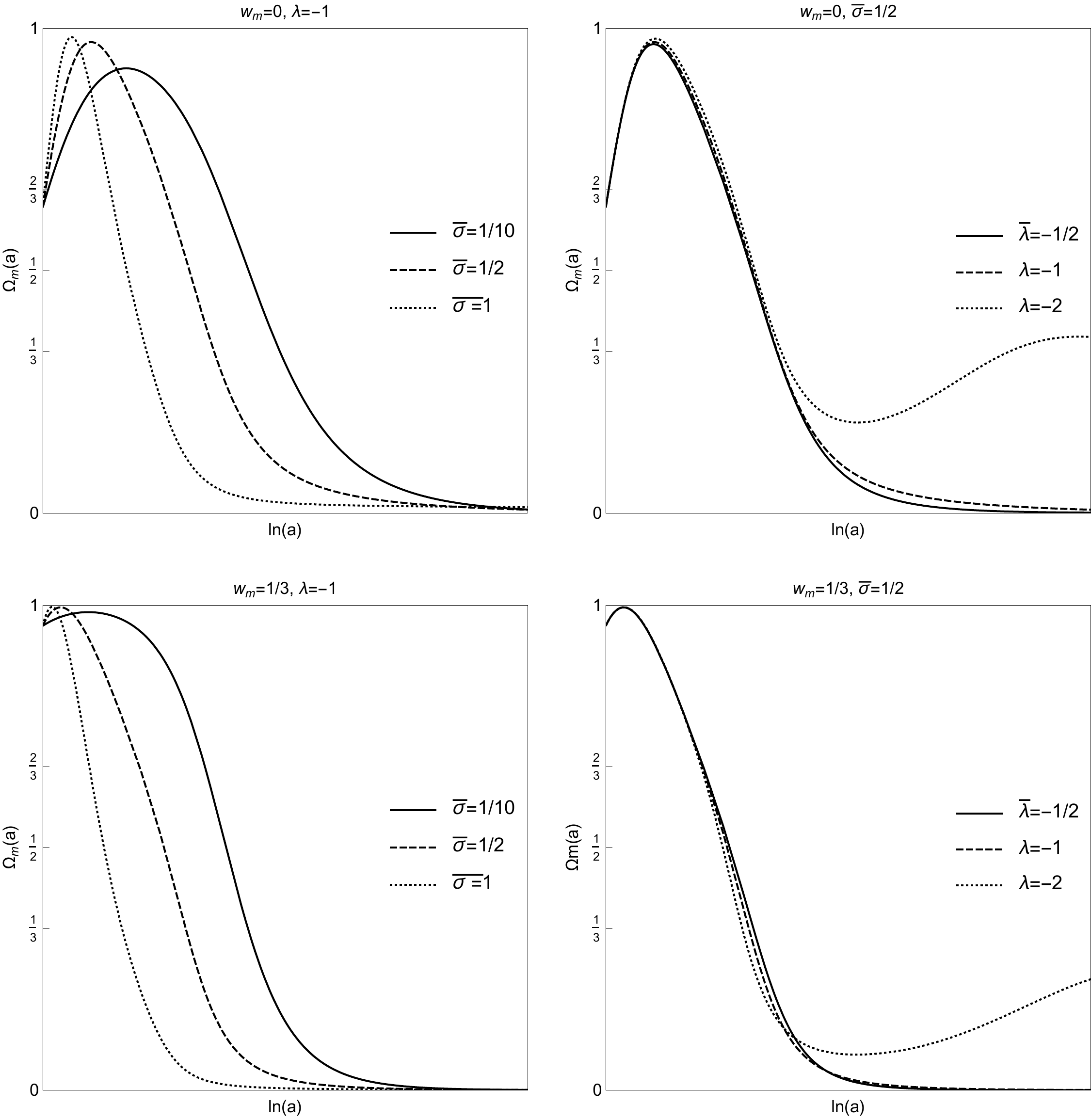}\caption{Qualitative
evolution for the $\Omega_{m}\left(  a\right)  $ as it is given by the
numerical solution of the dynamical system (\ref{de.07}), (\ref{de.08}) of
Model A for different values of the free parameters and the same initial
conditions with Fig. \ref{mfig3}.}%
\label{mfig3o}%
\end{figure}

Over and above in Fig. \ref{mfig2} we present the phase-space portrait for the
dynamical system (\ref{de.07}), (\ref{de.08}) for different values of the free
parameters such that all of the points appear as attractors in the different
plots. The qualitative evolution of the effective equation of state parameter
$w_{eff}$ is presented in Fig. \ref{mfig3}, while the evolution of the
$\Omega_{m}$ is presented in Fig. \ref{mfig3o}. From the latter figures we
observe that there exists a future attractor which describes an accelerated
universe. We have considered initial conditions for which the equation of
state parameter for the effective fluid is that of a radiation fluid, from
which we see that from the radiation epoch the universe goes to a solution
dominated by the matter source and ends to the acceleration solution.


In Table \ref{tab1} we summarize the stationary points and their physical
properties for Model A.%

\begin{table}[tbp] \centering
\caption{Stationary points and physical properties for Model A}%
\begin{tabular}
[c]{ccccc}\hline\hline
\textbf{Point} & $\left(  \mathbf{x,y}\right)  $ & $\Omega_{m}$ &
$\mathbf{w}_{eff}$ & \textbf{Can be Attractor?}\\\hline
$A_{1}^{\pm}$ & $\left(  \pm1,0\right)  $ & $0$ & $1$ & Yes\\
$A_{2}$ & $\left(  -\frac{\lambda}{\sqrt{6}},\sqrt{1-\frac{\lambda^{2}}{6}%
}\right)  $ & $0$ & $-1+\frac{\lambda^{2}}{3}$ & Yes\\
$A_{3}$ & $\left(  -\sqrt{\frac{2}{3}}\frac{1+\alpha}{1-w_{m}}\sigma,0\right)
$ & $1-\frac{2\bar{\sigma}^{2}}{3\left(  1-w_{m}\right)  ^{2}}$ & $\frac
{2\bar{\sigma}^{2}}{3\left(  1-w_{m}\right)  }+w_{m}$ & Yes\\
$A_{4}$ & $\left(  -\sqrt{\frac{3}{2}}\frac{1+w_{m}}{\lambda-\bar{\sigma}%
},\frac{\sqrt{\frac{3\left(  w_{m}^{2}-1\right)  }{\bar{\sigma}-\lambda}%
+2\bar{\sigma}}}{\sqrt{2\lambda-2\bar{\sigma}}}\right)  $ & $\frac
{\lambda\left(  \bar{\sigma}-\lambda_{0}\right)  -3\left(  1+w_{m}\right)
}{\left(  \lambda-\bar{\sigma}\right)  ^{2}}$ & $\frac{\bar{\sigma}%
+w_{m}\lambda}{\lambda-\bar{\sigma}}$ & Yes Always\\\hline\hline
\end{tabular}
\label{tab1}%
\end{table}%

\section{Model B: $V\left(  \phi\right)  =V_{0}e^{\lambda_{0}\phi}+\Lambda$
and $f\left(  \phi\right)  =f_{0}e^{\sigma_{0}\phi}$}

\label{sec5}

For the second model of our consideration we consider the scalar field
potential $V\left(  \phi\right)  =V_{0}e^{\lambda_{0}\phi}+\Lambda$ and the
coupling functions $f\left(  \phi\right)  =f_{0}e^{\sigma_{0}\phi}.$ We
calculate $\sigma=\sigma_{0},$ which means that $\sigma$ is always a constant,
but now $\lambda$ is a varying parameter defined as $\lambda=\frac
{V_{0}\lambda_{0}e^{\lambda_{0}\phi}}{V_{0}\lambda_{0}e^{\lambda_{0}\phi
}+\Lambda}$, that is, $\phi=\frac{1}{\lambda_{0}}\ln\left(  \frac
{\lambda\Lambda}{V_{0}\left(  \lambda_{0}-\lambda\right)  }\right)  $. This
transformation is valid for $\frac{\lambda\Lambda}{V_{0}\left(  \lambda
_{0}-\lambda\right)  }>0$. Moreover, we calculate $\Gamma\left(
\lambda\right)  =\frac{\lambda_{0}}{\lambda}$.

We end with the three-dimensional system%
\begin{equation}
\frac{dx}{d\tau}=\frac{1}{2}\left(  3x\left(  x^{2}-y^{2}-1+w_{m}\left(
1-x^{2}-y^{2}\right)  \right)  -\sqrt{6}\left(  \lambda y^{2}+\left(
1+\alpha\right)  \sigma\left(  1-x^{2}-y^{2}\right)  \right)  \right)  ,
\label{de.10}%
\end{equation}%
\begin{equation}
\frac{dy}{d\tau}=\frac{1}{2}y\left(  3+\sqrt{6}\lambda x+3\left(  x^{2}%
-y^{2}+w_{m}\left(  1-x^{2}-y^{2}\right)  \right)  \right)  , \label{de.11}%
\end{equation}%
\begin{equation}
\frac{d\lambda}{d\tau}=\sqrt{6}x\lambda\left(  \lambda_{0}-\lambda\right)  .
\label{de.12}%
\end{equation}

\begin{figure}[ptb]
\centering\includegraphics[width=0.5\textwidth]{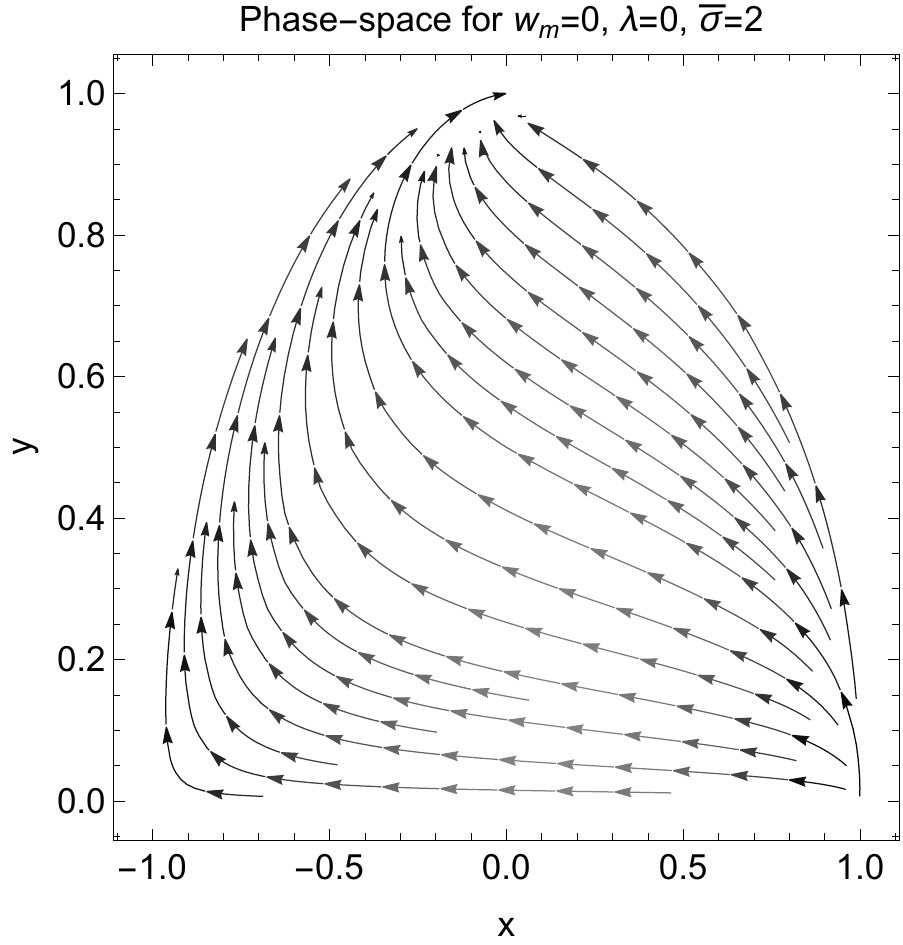}\caption{Phase-space
portrait of the dynamical system (\ref{de.10}), (\ref{de.11}) and
(\ref{de.12}) of Model B on the two-dimensional surface $\lambda=0$, for
$w_{m}=0$ from which it follows that the de Sitter universe is a future
attractor for the dynamical system. }%
\label{mfig5}%
\end{figure}

The stationary points $B=\left(  x\left(  B\right)  ,y\left(  B\right)
,\lambda\left(  B\right)  \right)  $ of this dynamical system and their
physical properties are discussed in the following lines.%

\[
B_{1}^{\pm}=\left(  \pm1,0,\lambda_{0}\right)  ,
\]
they have similar physical properties with that of $A_{1}^{\pm}$, that is,
$\Omega_{m}\left(  B_{1}^{\pm}\right)  =0$ and $w_{eff}\left(  B_{1}^{\pm
}\right)  =1$. As far as the stability properties of the points are concerned,
we derive the eigenvalues for the linearised system $e_{1}\left(  B_{1}^{\pm
}\right)  =\frac{\left(  6+\sqrt{6}\lambda_{0}\right)  }{2}$,~$e_{2}\left(
B_{1}^{\pm}\right)  =\mp\sqrt{6}\lambda_{0}$ and $e_{3}\left(  B_{1}^{\pm
}\right)  =3\left(  1-w_{m}\right)  \pm\sqrt{6}\bar{\sigma}$, with
$\bar{\sigma}=\left(  1+\alpha\right)  \sigma$; from which we infer that the
stationary points cannot describe stable solutions. Specifically for
$-\sqrt{6}<\lambda<0$ and $-\sqrt{6}\bar{\sigma}<3\left(  1-w_{m}\right)  $
point $B_{1}^{+}$ is a source, while for $0<\lambda<\sqrt{6}$ and $\sqrt
{6}\bar{\sigma}<3\left(  1-w_{m}\right)  $ point $B_{1}^{-}$ is a source. For
other values of the free parameters points $B_{1}^{\pm}$ are saddle points.
\[
B_{2}=\left(  -\frac{\lambda_{0}}{6},\sqrt{1-\frac{\lambda_{0}^{2}}{6}%
},\lambda_{0}\right)
\]
is the extension of point $A_{2}$ in the three-dimensional space.\ Therefore
the physical properties and the existence conditions are the same. The
eigenvalues are derived to be $e_{1}\left(  B_{2}\right)  =\frac{\left(
\lambda_{0}^{2}-6\right)  }{2}$, $e_{2}\left(  B_{2}\right)  =-3\left(
1+w_{m}\right)  +\lambda_{0}\left(  \lambda_{0}-\bar{\sigma}\right)  $ and
$e_{3}\left(  B_{2}\right)  =\lambda_{0}^{2}$. Therefore, for $\lambda_{0}%
^{2}>6$ and $\lambda_{0}\bar{\sigma}<-3\left(  1+w_{m}\right)  +\lambda
_{0}^{2}$ the stationary point is a source. Otherwise it is a saddle point.%

\[
B_{3}=\left(  -\sqrt{\frac{2}{3}}\frac{\bar{\sigma}}{1-w_{m}},0,\lambda
_{0}\right)  ,
\]
has the same physical properties as point $A_{3}$. The corresponding
eigenvalues are $e_{1}\left(  B_{3}\right)  =-\frac{3}{2}\left(
1-w_{m}\right)  +\frac{\bar{\sigma}^{2}}{1-w_{m}}$ and $e_{2}\left(
B_{3}\right)  =\frac{3}{2}\left(  1+w_{m}\right)  +\frac{\bar{\sigma}\left(
\bar{\sigma}-\lambda\right)  }{1-w_{m}}$ and $e_{3}\left(  B_{3}\right)
=\frac{2\lambda_{0}\bar{\sigma}}{1-w_{m}}$. We conclude that the stationary
point, when it exists, is always a saddle point.%

\[
B_{4}=\left(  -\sqrt{\frac{3}{2}}\frac{1+w_{m}}{\lambda_{0}-\bar{\sigma}%
},\frac{\sqrt{\frac{3\left(  w_{m}^{2}-1\right)  }{\bar{\sigma}-\lambda_{0}%
}+2\bar{\sigma}}}{\sqrt{2\lambda_{0}-2\bar{\sigma}}},\lambda_{0}\right)  ,
\]
with asymptotic solution $\Omega_{m}\left(  B_{4}\right)  =\frac
{\lambda\left(  \bar{\sigma}-\lambda_{0}\right)  -3\left(  1+w_{m}\right)
}{\left(  \lambda-\bar{\sigma}\right)  ^{2}}$ and $w_{eff}\left(
A_{4}\right)  =\frac{\bar{\sigma}+w_{m}\lambda_{0}}{\lambda-\bar{\sigma}}$ and
eigenvalues $e_{1,2}\left(  B_{4}\right)  =e_{1,2}\left(  A_{4}\right)  $ and
$e_{3}=\frac{3\left(  1-w_{m}\right)  \lambda_{0}}{\lambda_{0}-\bar{\sigma}}$.
The stability conditions are similar to those for the point $A_{4}$. Indeed,
when point $B_{4}$ exists, it is is always an attractor.
\[
B_{5}=\left(  0,1,0\right)  ,
\]
describes a de Sitter universe with $\Omega_{m}\left(  B_{5}\right)  =0$ and
$w_{eff}\left(  B_{5}\right)  =-1$. The eigenvalues are $e_{1}\left(
B_{5}\right)  =-3$, $e_{2}\left(  B_{5}\right)  =-3\left(  1+w_{m}\right)  $
and $e_{3}\left(  B_{5}\right)  =0$. Because of the zero eigenvalue we cannot
infer about the stability of the dynamical system. However, we know that it
may have a stable submanifold. Indeed on the surface $\lambda_{0}=0$, the
stationary point is always an attractor. That is observed in Fig. \ref{mfig5}.
The analytic submanifold may be derived with the use of the center manifold
theorem, but for simplicity we omit such analysis. However for equation
(\ref{de.12}) and $\lambda=\varepsilon\lambda_{\varepsilon}$, $\varepsilon
^{2}\rightarrow0$, it follows $\frac{d\lambda_{\varepsilon}}{d\tau}=\sqrt
{6}x\lambda_{0}\lambda_{\varepsilon}.$ Therefore for $x\lambda_{0}>0$,
$\lambda_{\varepsilon}$ increases. Otherwise for $x\lambda_{0}<0$,
$\lambda_{\varepsilon}$ decays.

\begin{figure}[ptb]
\centering\includegraphics[width=1\textwidth]{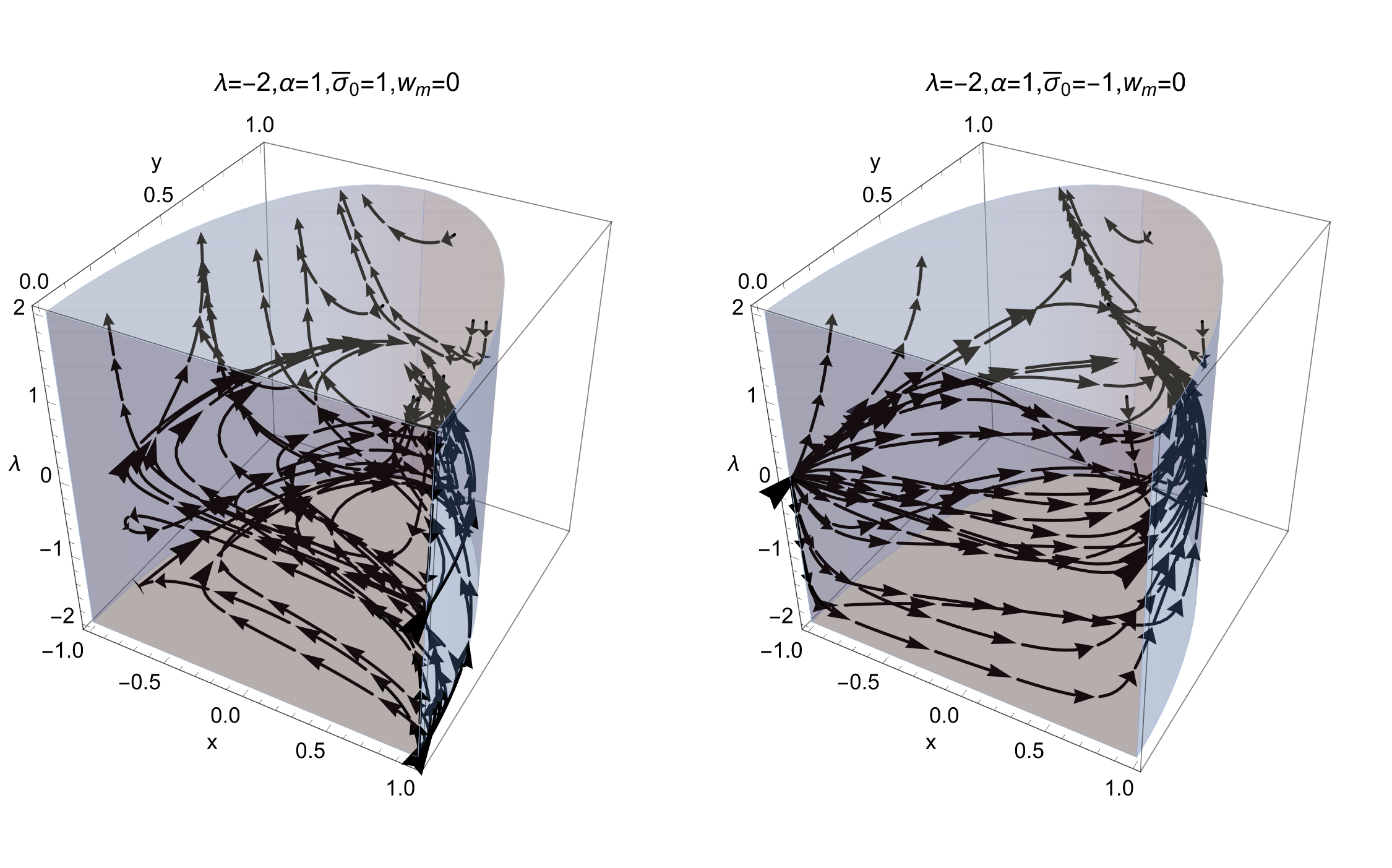}\caption{Phase-space
portraits of the dynamical system (\ref{de.10}), (\ref{de.11}) and
(\ref{de.12}) of Model B for $w_{m}=0$ and $\lambda_{0}=-1$, $\bar{\sigma}=2$
(left fig.) and $\bar{\sigma}=-2$ (right fig.) From the plots we observe that
for initial conditions with $\lambda<0$, the de Sitter universe is a future
attractor. }%
\label{mfig4}%
\end{figure}

Furthermore, in Fig. \ref{mfig4} we present the phase-space portraits for the
three-dimensional dynamical system (\ref{de.10}), (\ref{de.11}) and
(\ref{de.12}) for $\lambda_{0}=-1$, $w_{m}=0$ and $\bar{\sigma}^{2}=2$. We
conclude that for initial conditions with $\lambda\leq0$ the de Sitter
universe described by $B_{5}$ is a future attractor in the surface
$x\lambda_{0}<0$. 

\subsection{Analysis at Infinity}

For the dynamical system (\ref{de.10}), (\ref{de.11}) and (\ref{de.12})
variables $x$ and $y$ are constrained by the algebraic condition $x^{2}%
+y^{2}\leq1$, but that is not true for the dynamical variable $\lambda$ which
can take values at infinity, as can be observed by Fig. \ref{mfig4}. In order
to investigate the existence of stationary points at the infinity we define
the new variable $\lambda=\frac{\lambda_{b}}{\sqrt{1-\lambda_{b}^{2}}}$, from
which it follows that infinity is reached when $\lambda_{b}^{2}=1$. We define
the new independent variable $d\tau=\sqrt{1-\lambda_{b}^{2}}d\bar{\tau}$ ,
that is, at infinity the dynamical system (\ref{de.10}), (\ref{de.11}) and
(\ref{de.12}) becomes%
\begin{equation}
\frac{dx}{d\bar{\tau}}=\mp\frac{\sqrt{6}}{2}y^{2}~,~\frac{dy}{d\bar{\tau}}%
=\pm\frac{\sqrt{6}}{2}xy~\text{,}~\frac{d\lambda_{b}}{d\bar{\tau}}=\sqrt
{6}x,~\text{for }\lambda_{b}\rightarrow\pm1. \label{s1}%
\end{equation}

Therefore, at infinity there exists the unique stationary point $B_{\inf
}=\left(  0,0,\lambda_{b}\rightarrow\pm1\right)  $, which describes a matter
dominated universe, $\Omega_{m}\left(  B_{\inf}\right)  =1$ and $w_{eff}%
\left(  B_{\inf}\right)  =w_{m}.$ From (\ref{s1}) it is easy to see that the
stationary point at infinity is a homothetic center in the surface
$\lambda_{b}^{2}=1$. However, for $x>0$, $\frac{d\lambda_{b}}{d\bar{\tau}}>0$
while for $x<0$, $\frac{d\lambda_{b}}{d\bar{\tau}}<0$, from which we can
easily infer that point $B_{\inf}$ is always a saddle point.

In Table \ref{tab2} we summarize the results for Model B.%

\begin{table}[tbp] \centering
\caption{Stationary points and physical properties for Model B}%
\begin{tabular}
[c]{ccccc}\hline\hline
\textbf{Point} & $\left(  \mathbf{x,y,\lambda}\right)  $ & $\Omega_{m}$ &
$\mathbf{w}_{eff}$ & \textbf{Can be Attractor?}\\\hline
$B_{1}^{\pm}$ & $\left(  \pm1,0,\lambda_{0}\right)  $ & $0$ & $1$ & No\\
$B_{2}$ & $\left(  -\frac{\lambda}{\sqrt{6}},\sqrt{1-\frac{\lambda^{2}}{6}%
},\lambda_{0}\right)  $ & $0$ & $-1+\frac{\lambda_{0}^{2}}{3}$ & No\\
$B_{3}$ & $\left(  -\sqrt{\frac{2}{3}}\frac{1+\alpha}{1-w_{m}}\sigma
,0,\lambda_{0}\right)  $ & $1-\frac{2\bar{\sigma}^{2}}{3\left(  1-w_{m}%
\right)  ^{2}}$ & $\frac{2\bar{\sigma}^{2}}{3\left(  1-w_{m}\right)  }+w_{m}$
& No\\
$B_{4}$ & $\left(  -\sqrt{\frac{3}{2}}\frac{1+w_{m}}{\lambda-\bar{\sigma}%
},\frac{\sqrt{\frac{3\left(  w_{m}^{2}-1\right)  }{\bar{\sigma}-\lambda}%
+2\bar{\sigma}}}{\sqrt{2\lambda-2\bar{\sigma}}},\lambda_{0}\right)  $ &
$\frac{\lambda\left(  \bar{\sigma}-\lambda_{0}\right)  -3\left(
1+w_{m}\right)  }{\left(  \lambda-\bar{\sigma}\right)  ^{2}}$ & $\frac
{\bar{\sigma}+w_{m}\lambda}{\lambda_{0}-\bar{\sigma}}$ & Yes Always\\
$B_{5}$ & $\left(  0,1,0\right)  $ & $0$ & $-1$ & Yes\\
$B_{\inf}$ & $\left(  0,0,\lambda\rightarrow\pm\infty\right)  $ & $1$ &
$w_{m}$ & No\\\hline\hline
\end{tabular}
\label{tab2}%
\end{table}%

\section{Model C: $V\left(  \phi\right)  =V_{0}e^{\lambda\phi}$ and $f\left(
\phi\right)  =f_{0}e^{\sigma_{0}\phi}+\Lambda$}

\label{sec6}

For the third model we assume that the scalar field potential and the coupling
functions are $V\left(  \phi\right)  =V_{0}e^{\lambda\phi}$ and $f\left(
\phi\right)  =V_{0}e^{\sigma_{0}\phi}+\Lambda$. Therefore, we have the
three-dimensional dynamical system%
\begin{equation}
\frac{dx}{d\tau}=\frac{1}{2}\left(  3x\left(  x^{2}-y^{2}-1+w_{m}\left(
1-x^{2}-y^{2}\right)  \right)  -\sqrt{6}\left(  \lambda y^{2}+\left(
1+\alpha\right)  \sigma\left(  1-x^{2}-y^{2}\right)  \right)  \right)  ,
\label{de.20}%
\end{equation}%
\begin{equation}
\frac{dy}{d\tau}=\frac{1}{2}y\left(  3+\sqrt{6}\lambda x+3\left(  x^{2}%
-y^{2}+w_{m}\left(  1-x^{2}-y^{2}\right)  \right)  \right)  , \label{de.21}%
\end{equation}%
\begin{equation}
\frac{d\sigma}{d\tau}=\sqrt{6}x\sigma\left(  \sigma_{0}-\sigma\right)  ,
\label{de.22}%
\end{equation}
where $\lambda$ is a constant parameter. The stationary points $C=\left(
x\left(  C\right)  ,y\left(  C\right)  ,\sigma\left(  C\right)  \right)  $ of
this dynamical system are given below.%

\[
C_{1}^{\pm}=\left(  \pm1,0,0\right)  ,
\]
with physical properties similar to those of points $A_{1}^{\pm}$. The
eigenvalues of the linearised system are $e_{1}\left(  C_{1}^{\pm}\right)
=3\pm\sqrt{\frac{3}{2}}\lambda$, $e_{2}\left(  C_{1}^{\pm}\right)  =3\left(
1-w_{m}\right)  \pm\sqrt{6}\bar{\sigma}_{0}$ and $e_{3}\left(  C_{1}^{\pm
}\right)  =\mp\frac{\sqrt{6}\bar{\sigma}_{0}}{\left(  1+\alpha\right)  }$; in
which $\bar{\sigma}_{0}=\sigma_{0}\left(  1+\alpha\right)  $ and $\bar{\sigma
}=\sigma\left(  1+\alpha\right)  $. Therefore, $A_{1}^{+}$ is an attractor for
$\lambda<-\sqrt{6}$, $\bar{\sigma}_{0}>\frac{3}{\sqrt{6}}\left(
1-w_{m}\right)  $ and $\frac{\sqrt{6}\bar{\sigma}_{0}}{\left(  1+\alpha
\right)  }>0$. On the other hand point $A_{1}^{-}$ is an attractor for
$\lambda>\sqrt{6}$, $\bar{\sigma}_{0}<\frac{3}{\sqrt{6}}\left(  1-w_{m}%
\right)  $ and $\frac{\sqrt{6}\bar{\sigma}_{0}}{\left(  1+\alpha\right)  }<0$.%

\[
C_{2}=\left(  -\frac{\lambda}{\sqrt{6}},\sqrt{1-\frac{\lambda^{2}}{6}}%
,\bar{\sigma}_{0}\right)  ,
\]
with $\Omega\left(  C_{2}\right)  =0$ and $w_{eff}\left(  C_{2}\right)
=1+\frac{\lambda^{2}}{3}$. Indeed, point $C_{2}$ is the extension of $A_{2}$
in the three-dimensional manifold of variables $\left\{  x,y,\sigma\right\}
$. We calculate the eigenvalues $e_{1}\left(  C_{2}\right)  =\frac{\left(
\lambda^{2}-6\right)  }{2}$, $e_{2}\left(  C_{2}\right)  =-3\left(
1+w_{m}\right)  +\lambda\left(  \lambda-\sigma_{0}\right)  $ and $e_{3}\left(
C_{2}\right)  =\frac{\lambda\bar{\sigma}_{0}}{1+\alpha}$. Hence for
$\lambda^{2}<6$, $\frac{\lambda\bar{\sigma}_{0}}{1+\alpha}<0$ and $3\left(
1+w_{m}\right)  +\lambda\left(  \lambda-\sigma_{0}\right)  <0$ the stationary
point is an attractor.%

\[
C_{3}=\left(  -\sqrt{\frac{2}{3}}\frac{\bar{\sigma}_{0}}{1-w_{m}},0,\sigma
_{0}\right)  ,
\]
with physical variables $\Omega_{m}\left(  C_{3}\right)  =1-\frac{2\bar
{\sigma}_{0}^{2}}{3\left(  1-w_{m}\right)  ^{2}}$ and $w_{eff}\left(
C_{3}\right)  =\frac{2\bar{\sigma}_{0}^{2}}{3\left(  1-w_{m}\right)  }+w_{m}$.
The eigenvalues are calculated $e_{1}\left(  C_{1}\right)  =-\frac{3}%
{2}\left(  1-w_{m}\right)  +\frac{\bar{\sigma}_{0}^{2}}{1-w_{m}},~e_{2}\left(
C_{3}\right)  =\frac{3}{2}\left(  1+w_{m}\right)  -\frac{\bar{\sigma}%
_{0}\left(  \lambda-\bar{\sigma}_{0}\right)  }{\left(  1-w_{m}\right)  }$ and
$e_{3}\left(  C_{3}\right)  =\frac{2\left(  \bar{\sigma}_{0}\right)  ^{2}%
}{\left(  1-w_{m}\right)  \left(  1+\alpha\right)  }$. Therefore $C_{3}$
describes a stable solution and $C_{3}$ is an attractor for $\alpha<-1$ and
$\sqrt{6}\left\vert \sigma_{0}\right\vert <3\left(  1-w_{m}\right)  ,$
$\lambda+\frac{3\left(  w_{m}^{2}-1\right)  }{2\left\vert \bar{\sigma}%
_{0}\right\vert }<\left\vert \bar{\sigma}_{0}\right\vert $.%

\[
C_{4}=\left(  -\sqrt{\frac{3}{2}}\frac{1+w_{m}}{\lambda_{0}-\bar{\sigma}%
},\frac{\sqrt{\frac{3\left(  w_{m}^{2}-1\right)  }{\bar{\sigma}-\lambda_{0}%
}+2\bar{\sigma}}}{\sqrt{2\lambda_{0}-2\bar{\sigma}}},\sigma_{0}\right)  ,
\]
is the extension of point $A_{4}$ in the three-dimensional space. The
eigenvalues are $e_{1,2}\left(  C_{4}\right)  =e_{1,2}\left(  A_{4}\right)  $
and $e_{3}=\frac{3\left(  1+w_{m}\right)  \bar{\sigma}_{0}}{\left(
1+\alpha\right)  \left(  \lambda-\bar{\sigma}_{0}\right)  }$ from which we can
easily conclude that, when the point exist, is is always an attractor.
Similarly to the property of points $A_{4}$ and $B_{4}$.%

\[
C_{5}=\left(  0,0,0\right)
\]
describes a universe dominated by the ideal gas, that is, $\Omega_{m}\left(
C_{5}\right)  =1$ and $w_{eff}\left(  C_{5}\right)  =w_{m}.$ That is the new
point provided by the coupling function $f\left(  \phi\right)  =V_{0}%
e^{\sigma_{0}\phi}+\Lambda$. By definition $\phi=\frac{1}{\sigma_{0}}%
\ln\left(  \frac{\sigma\Lambda}{f_{0}\left(  \sigma-\sigma_{0}\right)
}\right)  $, that is, the limit $\sigma_{0}$ means that $\phi\rightarrow
-\infty$. Therefore, for $\sigma=0$, the coupling function becomes $f\left(
\phi\right)  \rightarrow\Lambda.$ The eigenvalues of the linearised system are
$e_{1}\left(  C_{5}\right)  =-\frac{3}{2}\left(  1+w_{m}\right)  $,
$e_{2}\left(  C_{5}\right)  =\frac{3}{2}\left(  1-w_{m}\right)  $ and
$e_{3}\left(  C_{5}\right)  =0$, from which we infer that point $C_{5}$
describes always an unstable asymptotic solution and $C_{5}$ is a saddle point.

\begin{figure}[ptb]
\centering\includegraphics[width=1\textwidth]{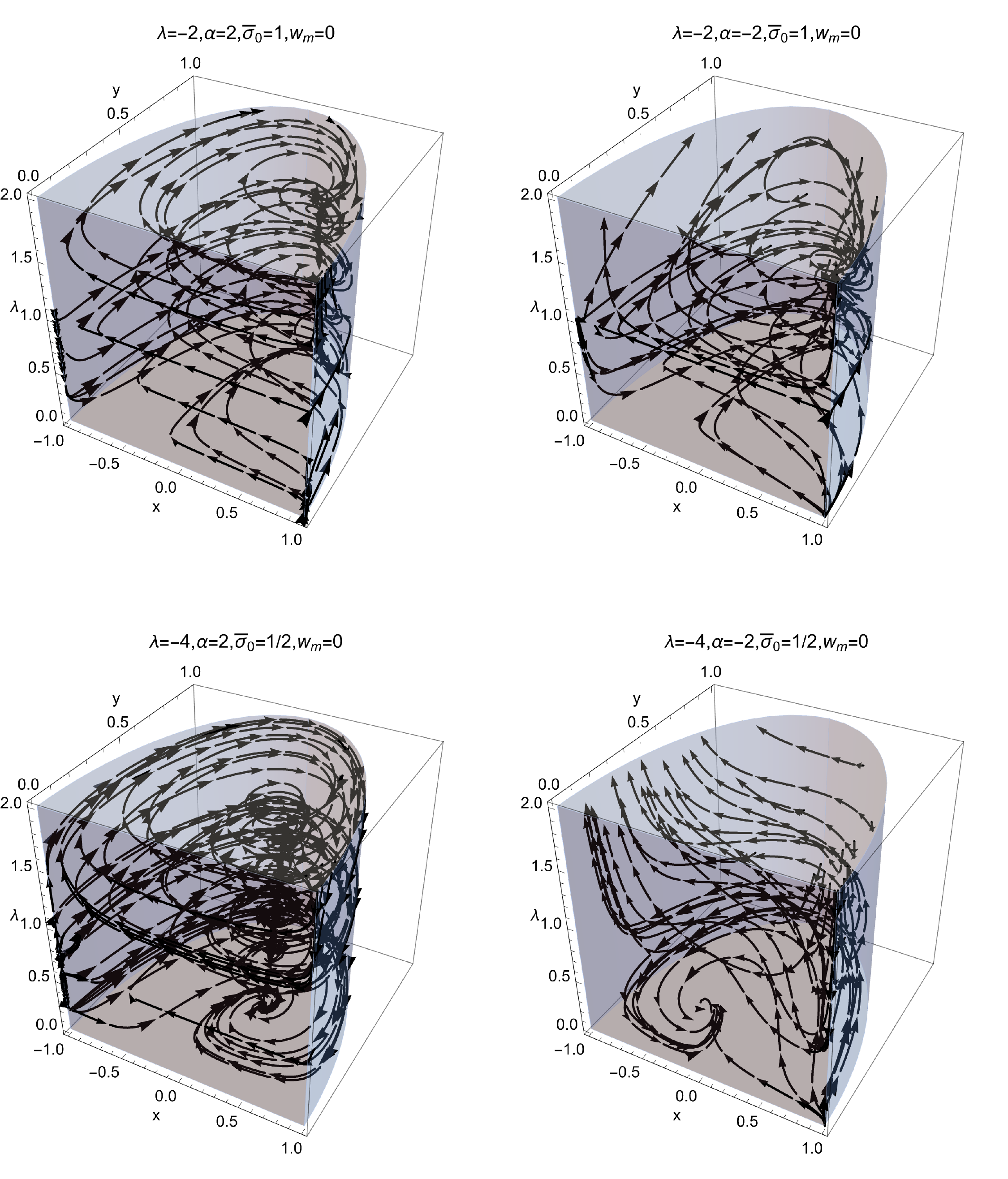}\caption{Phase-space
portraits of the dynamical system (\ref{de.20}), (\ref{de.21}) and
(\ref{de.22}) of Model C, for different values of the free parameters.}%
\label{mfig6}%
\end{figure}

In Fig. \ref{mfig6} we present three-dimensional phase-space portraits of the
dynamical system (\ref{de.20}), (\ref{de.21}) and (\ref{de.22}) for different
values of the free parameters.

\begin{figure}[ptb]
\centering\includegraphics[width=1\textwidth]{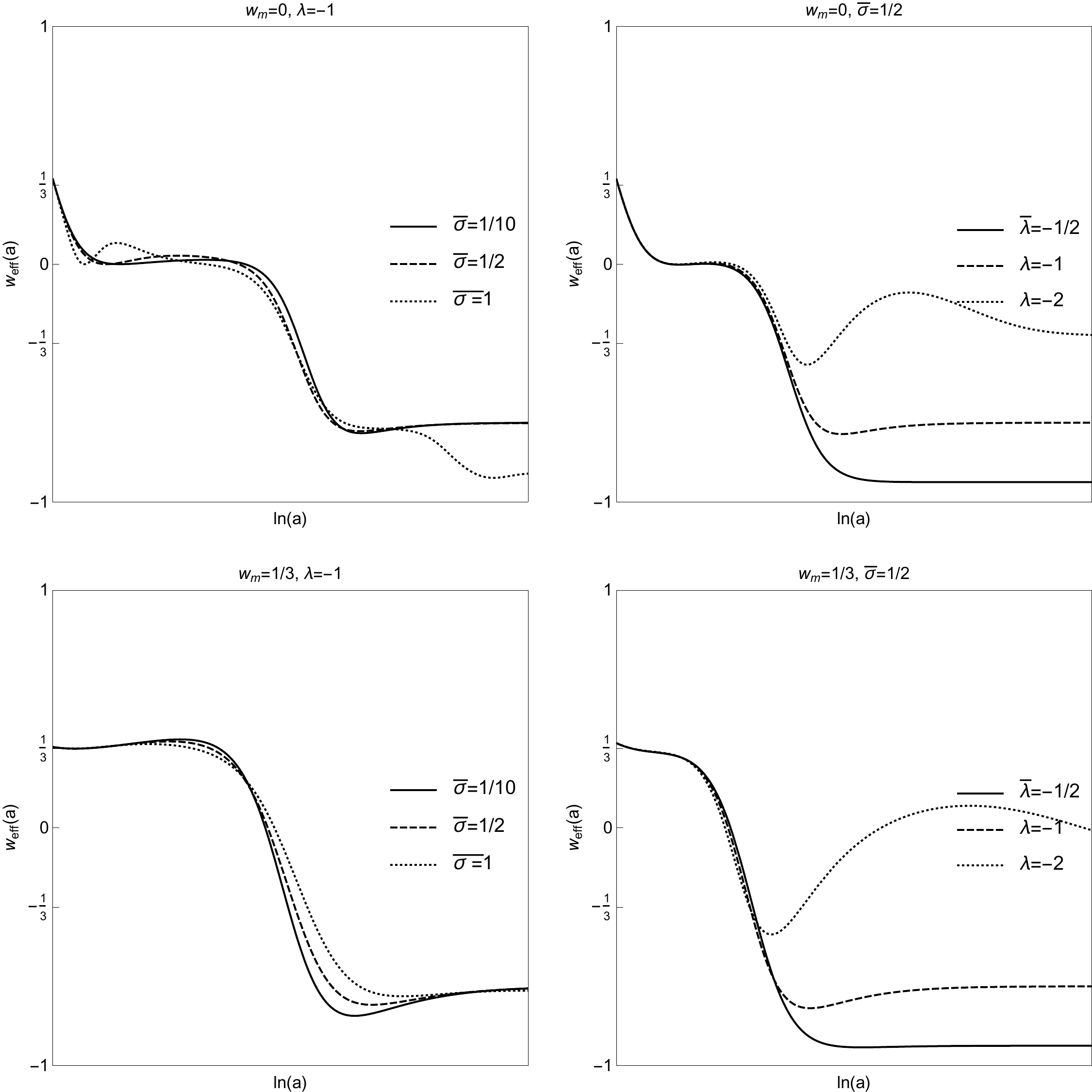}\caption{Qualitative
evolution for the effective equation of state parameter $w_{eff}\left(
a\right)  $ as it is given by the numerical solution of the dynamical system
(\ref{de.20}), (\ref{de.21}) and (\ref{de.22}) of Model C for different values
of the free parameters. }%
\label{mfig7}%
\end{figure}

\begin{figure}[ptb]
\centering\includegraphics[width=1\textwidth]{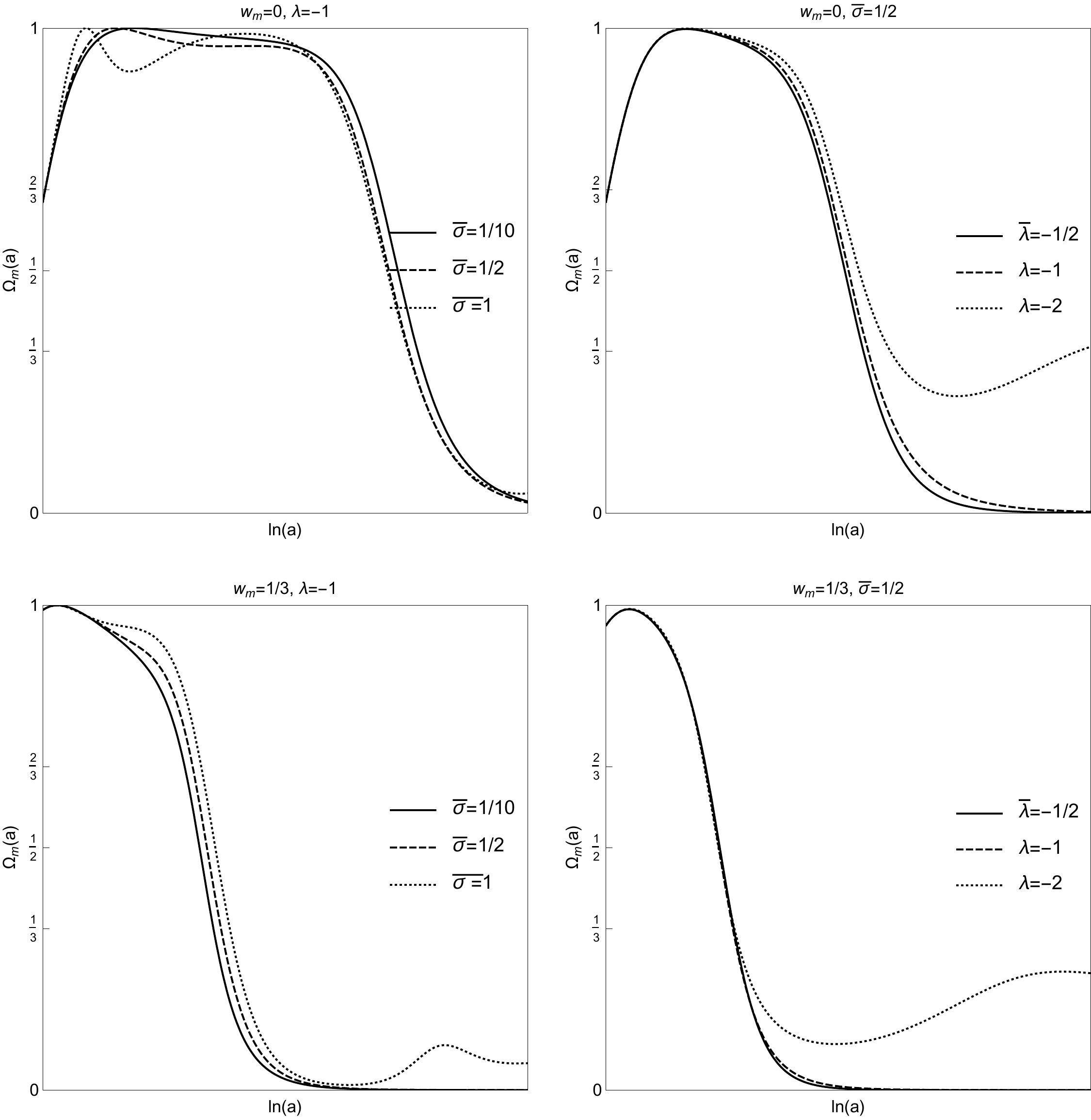}\caption{Qualitative
evolution for $\Omega\left(  a\right)  $ as it is given by the numerical
solution of the dynamical system (\ref{de.20}), (\ref{de.21}) and
(\ref{de.22}) of Model C for different values of the free parameters and the
same initial conditions with that of Fig. \ref{mfig7}}%
\label{mfig7o}%
\end{figure}

In Fig. \ref{mfig7} we present the qualitative evolution of the effective
equation of state for various values of the free parameters for this model. We
observe that for $w_{m}=0$, if we start for initial conditions near to the
radiation epoch, this model can reconstruct the cosmological history of the
late universe. Moreover, the evolution of $\Omega_{m}$ is presented in Fig.
\ref{mfig7o}. For the initial conditions of the latter figures we considered
the radiation epoch, that is, the effective fluid mimics the radiation fluid.
The trajectories go near to the saddle point which describes the matter era
and end at a universe dominated by the scalar field.\textbf{ }%


\subsection{Analysis at Infinity}

We apply the same procedure as for the Model B where now we consider the
variables $\sigma=\frac{\sigma_{b}}{\sqrt{1-\sigma_{b}^{2}}}$ and $d\tau
=\sqrt{1-\sigma_{b}^{2}}d\bar{\tau}$. At the limit of infinity the dynamical
system becomes%
\begin{equation}
\frac{dx}{d\bar{\tau}}=\pm\frac{\sqrt{6}}{2}\left(  1+\alpha\right)  \left(
x^{2}+y^{2}-1\right)  ,~\frac{dy}{d\bar{\tau}}=0\text{ , }\frac{d\sigma_{b}%
}{d\bar{\tau}}=\sqrt{6}x~\text{for }\sigma_{b}\rightarrow\pm1,
\end{equation}
from which it follows the stationary point $C_{\inf}=\left(  0,1,\sigma
_{b}^{2}\rightarrow1\right)  $. Point $C_{\inf}$ describes a de Sitter
universe with $\Omega_{m}\left(  C_{\inf}\right)  =0$ and $w_{eff}\left(
C_{\inf}\right)  =-1$. Similarly as before we conclude that the stationary
point describes an unstable solution. \ 

We summarize the results of this Section in Table \ref{tab3}.%

\begin{table}[tbp] \centering
\caption{Stationary points and physical properties for Model C}%
\begin{tabular}
[c]{ccccc}\hline\hline
\textbf{Point} & $\left(  \mathbf{x,y,\sigma}\right)  $ & $\Omega_{m}$ &
$\mathbf{w}_{eff}$ & \textbf{Can be Attractor?}\\\hline
$C_{1}^{\pm}$ & $\left(  \pm1,0,\sigma_{0}\right)  $ & $0$ & $1$ & Yes\\
$C_{2}$ & $\left(  -\frac{\lambda_{0}}{\sqrt{6}},\sqrt{1-\frac{\lambda_{0}%
^{2}}{6}},\sigma_{0}\right)  $ & $0$ & $-1+\frac{\lambda^{2}}{3}$ & Yes\\
$C_{3}$ & $\left(  -\sqrt{\frac{2}{3}}\frac{1+\alpha}{1-w_{m}}\sigma
,0,\sigma_{0}\right)  $ & $1-\frac{2\bar{\sigma}^{2}}{3\left(  1-w_{m}\right)
^{2}}$ & $\frac{2\bar{\sigma}^{2}}{3\left(  1-w_{m}\right)  }+w_{m}$ & Yes\\
$C_{4}$ & $\left(  -\sqrt{\frac{3}{2}}\frac{1+w_{m}}{\lambda_{0}-\bar{\sigma}%
},\frac{\sqrt{\frac{3\left(  w_{m}^{2}-1\right)  }{\bar{\sigma}-\lambda}%
+2\bar{\sigma}}}{\sqrt{2\lambda_{0}-2\bar{\sigma}}},\sigma_{0}\right)  $ &
$\frac{\lambda\left(  \bar{\sigma}-\lambda_{0}\right)  -3\left(
1+w_{m}\right)  }{\left(  \lambda-\bar{\sigma}\right)  ^{2}}$ & $\frac
{\bar{\sigma}_{0}+w_{m}\lambda}{\lambda-\bar{\sigma}_{0}}$ & Yes Always\\
$C_{5}$ & $\left(  0,0,0\right)  $ & $1$ & $w_{m}$ & No\\
$C_{\inf}$ & $\left(  0,1,\sigma\rightarrow\pm\infty\right)  $ & $0$ & $-1$ &
No\\\hline\hline
\end{tabular}
\label{tab3}%
\end{table}%

\section{Model D: $V\left(  \phi\right)  =V_{0}e^{\lambda_{0}\phi}+\Lambda$
and $f\left(  \phi\right)  =f_{0}\left(  V_{0}e^{\lambda_{0}\phi}%
+\Lambda\right)  ^{p}$}

\label{sec7}

Consider now the cosmological model with scalar field potential $V\left(
\phi\right)  =V_{0}e^{\lambda_{0}\phi}+\Lambda$ and coupling function for the
chameleon mechanism $f\left(  \phi\right)  =f_{0}\left(  V\left(  \phi\right)
\right)  ^{p}$. For these functions it follows that $\sigma\left(
\lambda\right)  =p\lambda$. Thus, the field equations are expressed by the
following three-dimensional system%
\begin{equation}
\frac{dx}{d\tau}=\frac{1}{2}\left(  3x\left(  x^{2}-y^{2}-1+w_{m}\left(
1-x^{2}-y^{2}\right)  \right)  -\sqrt{6}\left(  \lambda y^{2}+\lambda\bar
{p}\left(  1-x^{2}-y^{2}\right)  \right)  \right)  , \label{de.31}%
\end{equation}%
\begin{equation}
\frac{dy}{d\tau}=\frac{1}{2}y\left(  3+\sqrt{6}\lambda x+3\left(  x^{2}%
-y^{2}+w_{m}\left(  1-x^{2}-y^{2}\right)  \right)  \right)  , \label{de.32}%
\end{equation}%
\begin{equation}
\frac{d\lambda}{d\tau}=\sqrt{6}x\lambda\left(  \lambda_{0}-\lambda\right)  ,
\label{de.33}%
\end{equation}
where $\bar{p}=\left(  1+\alpha\right)  p$. In the following lines we derive
the stationary points of this system and we discuss their stability
properties.%
\[
D_{1}=\left(  \pm1,0,\lambda_{0}\right)  ,
\]
have similar physical properties as points $A_{1}^{\pm}$. We calculate the
eigenvalues $e_{1}\left(  D_{1}^{\pm}\right)  =\frac{\left(  6+\sqrt{6}%
\lambda_{0}\right)  }{2}$,~$e_{2}\left(  D_{1}^{\pm}\right)  =\mp\sqrt
{6}\lambda_{0}$ and $e_{3}\left(  D_{1}^{\pm}\right)  =3\left(  1-w_{m}%
\right)  \pm\sqrt{6}\bar{p}\lambda_{0}$, \ from which it follows that for
$\bar{p}\left\vert \lambda_{0}\right\vert <\frac{3\left(  1-w_{m}\right)
}{\sqrt{6}}$, point $D_{1}^{+}$ is a source when $-\sqrt{6}<\lambda<0$, while
point $D_{1}^{-}$ is a source for $0<\lambda<\sqrt{6}$. Otherwise the two
points are saddle points.%
\[
D_{2}=\left(  -\frac{\lambda_{0}}{6},\sqrt{1-\frac{\lambda_{0}^{2}}{6}%
},\lambda_{0}\right)  ,
\]
with eigenvalues $e_{1}\left(  D_{2}\right)  =\frac{\left(  \lambda_{0}%
^{2}-6\right)  }{2}$, $e_{2}\left(  D_{2}\right)  =-3\left(  1+w_{m}\right)
+\lambda_{0}\left(  \lambda_{0}-\bar{p}\lambda_{0}\right)  $ and $e_{3}\left(
D_{2}\right)  =\lambda_{0}^{2}$ describes an unstable scaling solution similar
to that of point $B_{2}$ with the same stability properties, by replacing
$\bar{\sigma}=\bar{p}\lambda_{0}$.
\[
D_{3}=\left(  -\sqrt{\frac{2}{3}}\frac{\bar{p}\lambda_{0}}{1-w_{m}}%
,0,\lambda_{0}\right) .
\]
it is point $B_{3}$ and it has the same physical and stability properties as
for $\bar{\sigma}=\bar{p}\lambda_{0}$.%

\[
D_{4}=\left(  -\sqrt{\frac{3}{2}}\frac{1+w_{m}}{\lambda_{0}-\bar{p}\lambda
_{0}},\frac{\sqrt{\frac{3\left(  w_{m}^{2}-1\right)  }{\bar{p}\lambda
_{0}-\lambda_{0}}+2\bar{p}\lambda_{0}}}{\sqrt{2\lambda_{0}-2\bar{p}\lambda
_{0}}},\lambda_{0}\right) .
\]
it is point $B_{4}$ for $\bar{\sigma}=\bar{p}\lambda_{0}$, and it has the same
physical and stability properties, that is, point $D_{4}$ is an attractor when
it exists.%

\[
D_{5}=\left(  0,0,0\right)  ,
\]
it is an asymptotic solution in which the matter source dominates in the
universe similarly to point $C_{5}$ with the same eigenvalues. Hence, point
$D_{5}$ is always a saddle point.%

\[
D_{6}=\left(  0,1,0\right)  ,
\]
describes a de Sitter universe similar to point $B_{5}$ for which the
cosmological constant term dominates in the scalar field potential. The
stability properties are similar with those of point $B_{5}$.

\begin{figure}[ptb]
\centering\includegraphics[width=1\textwidth]{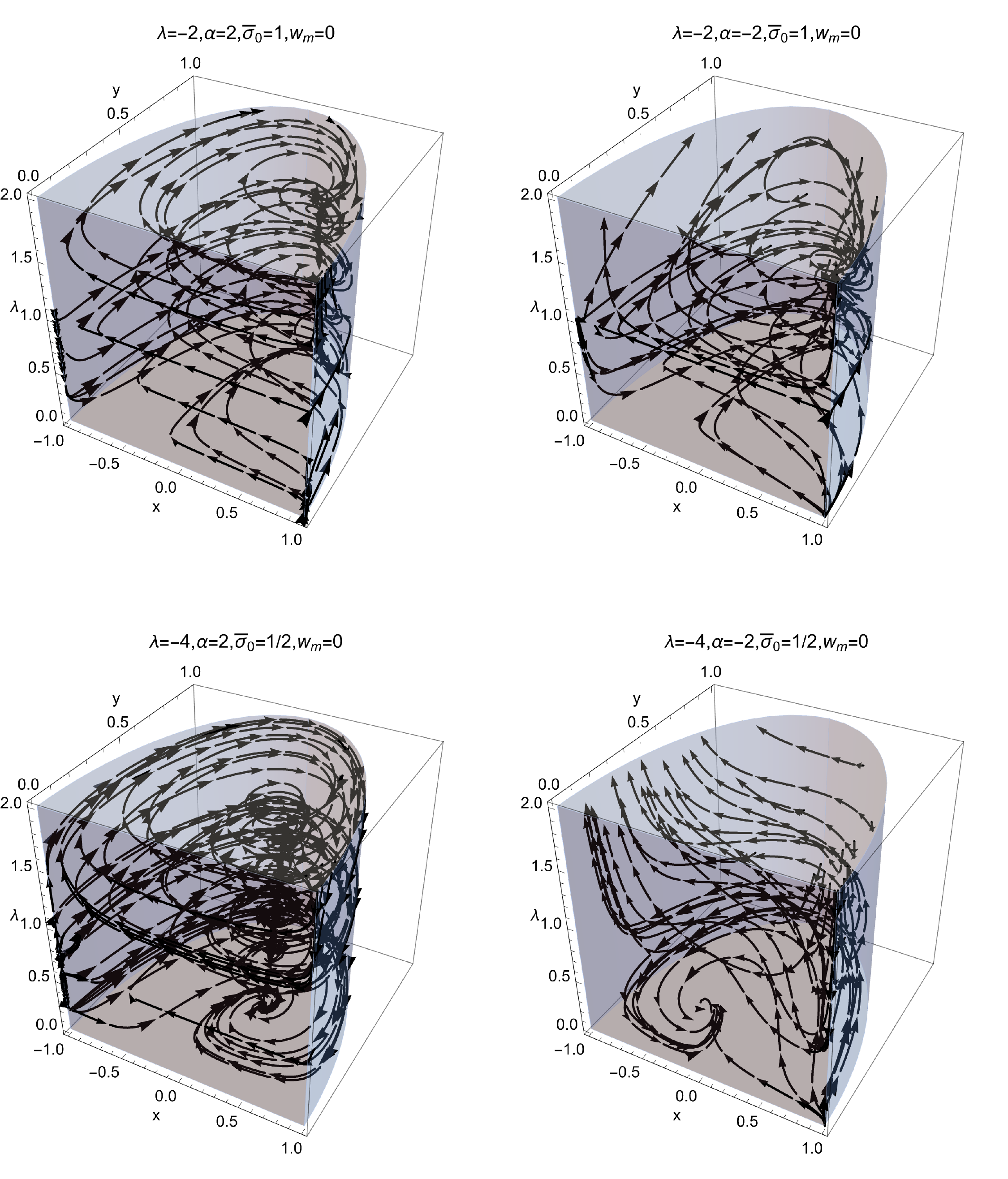}\caption{Phase-space
portraits of the dynamical system (\ref{de.31}), (\ref{de.32}) and
(\ref{de.33}) of Model D, for different values of the free parameters.}%
\label{mfig8}%
\end{figure}

\begin{figure}[ptb]
\centering\includegraphics[width=1\textwidth]{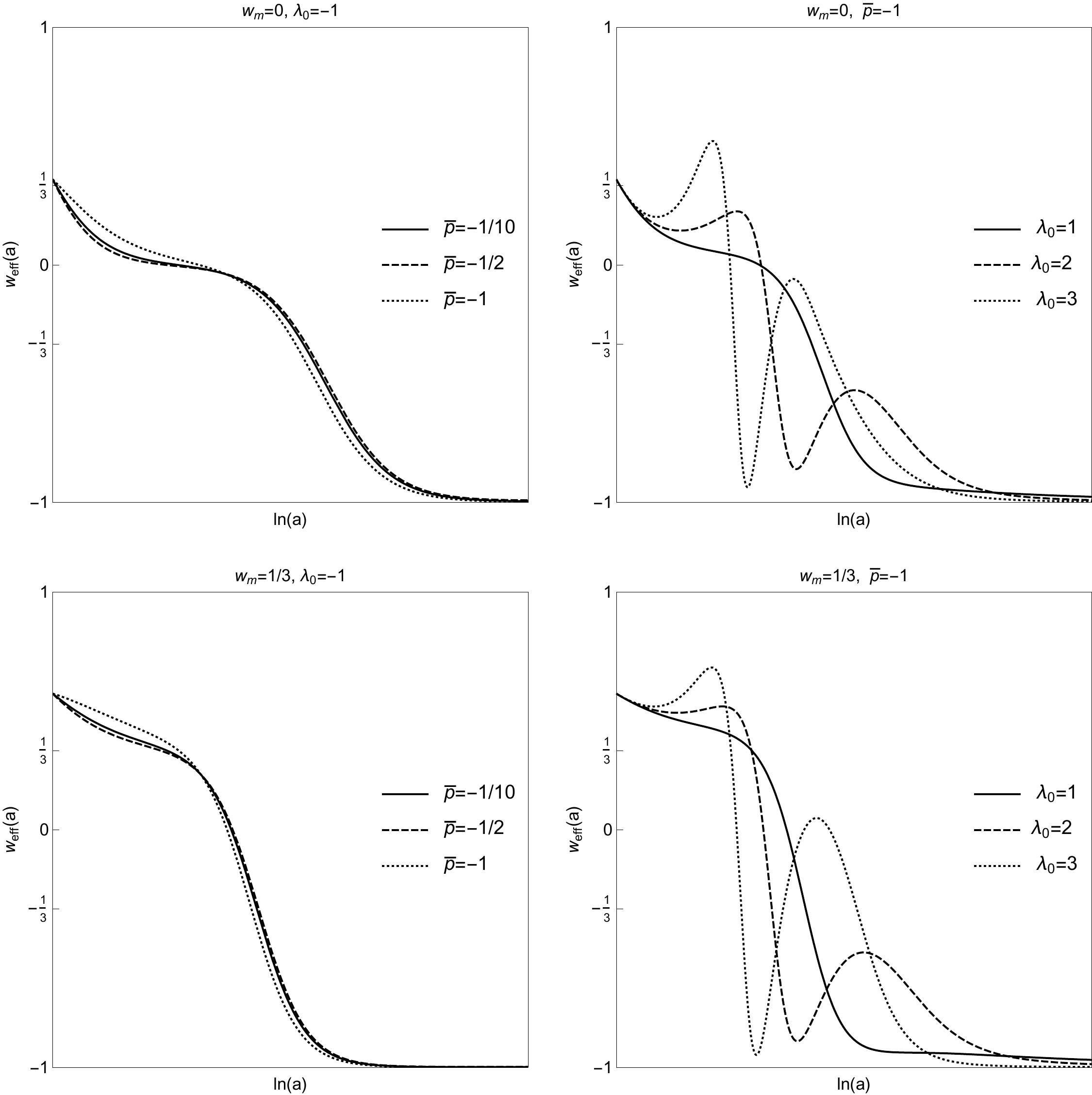}\caption{Qualitative
evolution for the effective equation of state parameter $w_{eff}\left(
a\right)  $ as it is given by the numerical solution of the dynamical system
(\ref{de.31}), (\ref{de.32}) and (\ref{de.33}) of model D for different values
of the free parameters. }%
\label{mfig9}%
\end{figure}

\begin{figure}[ptb]
\centering\includegraphics[width=1\textwidth]{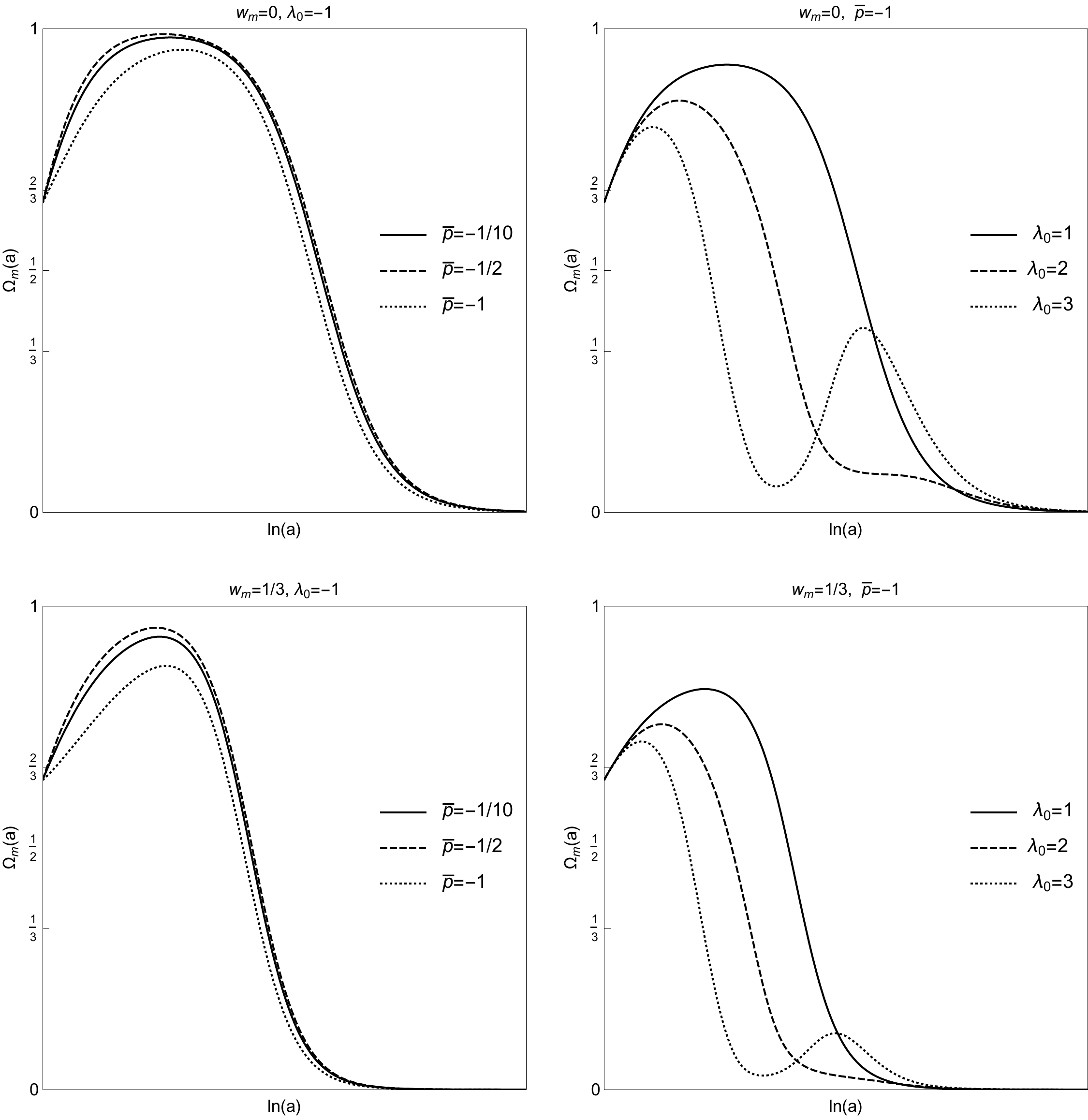}\caption{Qualitative
evolution for $\Omega_{m}\left(  a\right)  $ as given by the numerical
solution of the dynamical system (\ref{de.31}), (\ref{de.32}) and
(\ref{de.33}) of model D for different values of the free parameters and the
same initial conditions with that of Fig. \ref{mfig9}}%
\label{mfig9o}%
\end{figure}

In Fig. \ref{mfig8} we present the three-dimensional phase-space portrait for
the dynamical system (\ref{de.31}), (\ref{de.32}) and (\ref{de.33}) of Model
D. Furthermore, the qualitative evolution of the effective equation of state
parameter $w_{eff}~$is given in Fig. \ref{mfig9} and the dynamical evolution
of $\Omega_{m}$ is presented in Fig. \ref{mfig9o}. From this we have selected
a set of initial conditions whereby the dynamical system can describe the main
eras of the cosmological history. We observe that a main difference of the
evolution of $w_{eff}$ with that of Model C is that now the future attractor
can be a de Sitter universe, i.e. point $D_{6}$, instead of a scaling solution
described by point $C_{4}$.  For the initial conditions of the latter figures
we considered the effective fluid to describe radiation. From the evolution of
the trajectories we observe that the universe goes near to the saddle point
which describes the matter era and ends to a universe dominated by the scalar
field which is a future attractor. 

\subsection{Analysis at Infinity}

For the analysis at infinity we apply the same procedure as that for Model B.
At infinity, i.e. $\lambda_{b}^{2}\rightarrow1$, we end with the dynamical
system
\begin{equation}
\frac{dx}{d\bar{\tau}}=\mp\frac{\sqrt{6}}{2}\left(  \bar{p}\left(
1-x^{2}\right)  +\left(  1-\bar{p}\right)  y^{2}\right)  ~,~\frac{dy}%
{d\bar{\tau}}=\pm\frac{\sqrt{6}}{2}xy~\text{,}~\frac{d\lambda_{b}}{d\bar{\tau
}}=\sqrt{6}x~,
\end{equation}
which means that the unique stationary point for $\bar{p}\neq1$ is the
$D_{\inf}=\left(  0,\sqrt{\frac{\bar{p}}{\bar{p}-1}},\lambda_{b}\rightarrow
\pm1\right)  .$ For that stationary point we derive $\Omega\left(  D_{\inf
}\right)  =0$ and $w_{eff}\left(  D_{\inf}\right)  =-1$, which means that it
describes a de Sitter solution. On the surface $\lambda_{b}^{2}=1$ the point
is a homothetic center. However, from the differential equation for the
variable $\lambda_{b}$ it follows that $D_{\inf}$ is always a saddle point.
For $\bar{p}=1$, there exists a family of saddle stationary points $\bar
{D}_{\inf}=\left(  0,y,\lambda_{b}\rightarrow\pm1\right)  $ which describe
solutions with $\Omega\left(  \bar{D}_{\inf}\right)  =1-y^{2}\left(  \bar
{D}\right)  $ and$~w_{eff}\left(  \bar{D}\right)  =w_{m}\left(  1-y^{2}\left(
\bar{D}\right)  \right)  -y^{2}\left(  \bar{D}\right)  $.

That is a very interesting result because Model D can provide two de Sitter
epochs one unstable at infinity,in which can be related to inflation, and one
attractor related to the late-time acceleration phase.

The stability analysis for Model D is summarized in Table \ref{tab4}%

\begin{table}[tbp] \centering
\caption{Stationary points and physical properties for Model D}%
\begin{tabular}
[c]{ccccc}\hline\hline
\textbf{Point} & $\left(  \mathbf{x,y,\lambda}\right)  $ & $\Omega_{m}$ &
$\mathbf{w}_{eff}$ & \textbf{Can be Attractor?}\\\hline
$D_{1}^{\pm}$ & $\left(  \pm1,0,\lambda_{0}\right)  $ & $0$ & $1$ & No\\
$D_{2}$ & $\left(  -\frac{\lambda}{\sqrt{6}},\sqrt{1-\frac{\lambda^{2}}{6}%
},\lambda_{0}\right)  $ & $0$ & $-1+\frac{\lambda_{0}^{2}}{3}$ & No\\
$D_{3}$ & $\left(  -\sqrt{\frac{2}{3}}\frac{1+\alpha}{1-w_{m}}\sigma
,0,\lambda_{0}\right)  $ & $1-\frac{2\bar{\sigma}^{2}}{3\left(  1-w_{m}%
\right)  ^{2}}$ & $\frac{2\bar{\sigma}^{2}}{3\left(  1-w_{m}\right)  }+w_{m}$
& No\\
$D_{4}$ & $\left(  -\sqrt{\frac{3}{2}}\frac{1+w_{m}}{\lambda-\bar{\sigma}%
},\frac{\sqrt{\frac{3\left(  w_{m}^{2}-1\right)  }{\bar{\sigma}-\lambda}%
+2\bar{\sigma}}}{\sqrt{2\lambda-2\bar{\sigma}}},\lambda_{0}\right)  $ &
$1-\frac{2\bar{\sigma}^{2}}{3\left(  1-w_{m}\right)  ^{2}}$ & $\frac
{\bar{\sigma}+w_{m}\lambda}{\lambda_{0}-\bar{\sigma}}$ & Yes Always\\
$D_{5}$ & $\left(  0,0,0\right)  $ & $1$ & $0$ & No\\
$D_{6}$ & $\left(  0,1,0\right)  $ & $0$ & $-1$ & Yes\\
$D_{\inf}$ & $\left(  0,\sqrt{\frac{p}{p-1}},\lambda\rightarrow\pm
\infty\right)  $ & $0$ & $-1$ & No\\
$\bar{D}_{\inf}$ & $\left(  0,y^{2}\left(  \bar{D}\right)  ,\lambda
\rightarrow\pm\infty\right)  ~,~$for $\bar{p}=1$ & $1-y^{2}\left(  \bar
{D}\right)  $ & $w_{m}\left(  1-y^{2}\left(  \bar{D}\right)  \right)
-y^{2}\left(  \bar{D}\right)  $ & No\\\hline\hline
\end{tabular}
\label{tab4}%
\end{table}%

\section{Conclusions}

\label{sec8}

In this work we investigated the phase-space of the cosmological field
equations in chameleon dark energy. In particular we performed a detailed
dynamical analysis for the gravitational field equations in a spatially flat
FLRW with a scalar field coupled to an ideal gas. The coupling function is
responsible for the chameleon mechanism when the mass of the scalar field
depends upon the energy density of the ideal \ gas. In order to avoid the
violation of the weak energy condition, we assumed the scalar field to be that
of quintessence and the coupling function to be always of positive value. In
in such a scenario the Hubble function does not change sign. Consequently, for
the study of the dynamics we followed the $H$-normalization approach. We
defined a new set of dimensionless variables and we expressed all the physical
quantities in terms of these variables.

In the $H$-normalization approach the field equations are written in an
equivalent form of an algebraic-differential system where the independent
variable is the radius of the FLRW geometry, that is, the scale factor. With
the use of the algebraic equation the dimensional space of the field equation
is reduced to a maximum dimension of three. The dynamical evolution of the
physical variables it depends on the selection of two unknown functions, the
scalar field potential and the coupling function. We considered four different
sets for the free functions; for these four models we discussed in details the
evolution of the dynamics and the physical properties of the asymptotic solutions.

For the first cosmological model of our consideration, namely Model A, we
assumed that the potential and the coupling functions are exponential, that
is, $V\left(  \phi\right)  =V_{0}e^{\lambda_{0}\phi}$ and $f\left(
\phi\right)  =f_{0}e^{\sigma_{0}\phi}$. For the exponential coupling function
the theory is equivalent with the Weyl Integrable Spacetime. Furthermore, the
dimension of the field equations in the dimensional variables for this
selection of the unknown functions is reduced to two. The admitted stationary
points are four, the two points correspond to the quintessence model for which
the ideal gas does not contribute to the cosmological fluid, that is, we
derived the stiff fluid epoch in which the kinetic term of the scalar field
contributes in the field equations and a scaling solution which can describe
acceleration. On the other hand, for the remainder of the stationary points
the ideal gas contributes in the cosmological fluid. These points can describe
scaling solutions related to the matter or to the late-time acceleration of
the universe.

Model B was considered with $V\left(  \phi\right)  =V_{0}e^{\lambda_{0}\phi
}+\Lambda$ and $f\left(  \phi\right)  =f_{0}e^{\sigma_{0}\phi}$. In this case
the field equations form a three-dimensional system. On the two-dimensional
surface we recovered the four stationary points of Model A. Moreover we found
two new stationary points which describe a Sitter universe and a matter
dominated era. In order to determine the existence of the matter dominated era
we had to make use of Poincare' variables so as to investigate the asymptotic
dynamics at the infinity regime. In Model C with $V\left(  \phi\right)
=V_{0}e^{\lambda\phi}$ and $f\left(  \phi\right)  =f_{0}e^{\sigma_{0}\phi
}+\Lambda\,$, we determined the same stationary points as in Model B, but now
the de Sitter point appeared in the infinity regime while the matter dominated
solutions exist in the finite regime. It is important to mention that the
stability properties are different in each model.

Finally, for the fourth-model of our consideration with $V\left(  \phi\right)
=V_{0}e^{\lambda_{0}\phi}+\Lambda$ and $f\left(  \phi\right)  =f_{0}\left(
V_{0}e^{\lambda_{0}\phi}+\Lambda\right)  ^{p}~$with$~p\neq0$, we calculated
six stationary points at the finite regime which describe physical solutions
corresponding to the stationary points at the finite regime for the
cosmological Models A, B and C. Moreover, in the infinite regime for $p\neq1$,
an unstable de Sitter universe exists which can be related to the early
inflationary epoch of the universe, while for $p=1$, the unstable solution at
infinity describes a family of scaling solutions.

For each of the above mentioned cosmological models, we presented phase-space
portraits of the dynamical variables and the qualitative evolution of the
effective equation of state parameter for different sets of initial conditions
and values of the free parameters. The results from the stability analysis and
the phase-space portraits can be used to constrain the region of the free
variables for the initial condition problem. Furthermore, from the evolution
of the effective equation of state parameters we can conclude that these
models can reproduce the main eras of the cosmological history. Also, Model D
can be used to unify the all the components for the dark sector of the
universe and it shows that it can provide a mechanism to relate the dark
energy and the inflation responsible for the inflation.

In future work we plan to focus on the analysis of the perturbations and to
investigate if these models can solve cosmological tensions.


\begin{acknowledgments}
This work was partially financially supported in part by the National Research
Foundation of South Africa (Grant Numbers 131604). The author thanks the
support of Vicerrector\'{\i}a de Investigaci\'{o}n y Desarrollo
Tecnol\'{o}gico (Vridt) at Universidad Cat\'{o}lica del Norte through
N\'{u}cleo de Investigaci\'{o}n Geometr\'{\i}a Diferencial y Aplicaciones,
Resoluci\'{o}n Vridt No - 098/2022.
\end{acknowledgments}

\end{document}